\documentclass[useAMS,usenatbib]{mnras}
\input psfig.sty

\def \aj {AJ}
\def \apj {ApJ}
\def \apjl {ApJL}
\def \mnras {MNRAS}
\def \apjs {ApJS}
\def \aap {A\&A}
\def \nat {Natur}
\def \na {NewA}
\def \prd {PhysRevD}
\def \etal {et~al.~}

\def \spose#1{\hbox  to 0pt{#1\hss}}  
\def \lta{\mathrel{\spose{\lower 3pt\hbox{$\sim$}}\raise  2.0pt\hbox{$<$}}}
\def \gta{\mathrel{\spose{\lower  3pt\hbox{$\sim$}}\raise 2.0pt\hbox{$>$}}}

\def \kmsmpc {\>{\rm km}\,{\rm s}^{-1}\,{\rm Mpc}^{-1}}
\def \kms {\ifmmode  \,\rm km\,s^{-1} \else $\,\rm km\,s^{-1}  $ \fi }
\def \pc {\ifmmode  {\rm pc}  \else ${\rm pc}$ \fi  }
\def \kpc {\ifmmode  {\rm kpc}  \else ${\rm  kpc}$ \fi  }  
\def \Msun {\ifmmode {\rm M}_{\odot} \else M$_{\odot}$ \fi} 

\def \LCDM {\ifmmode \Lambda{\rm CDM} \else $\Lambda{\rm CDM}$ \fi}
\def \sig8 {\ifmmode \sigma_8 \else $\sigma_8$ \fi} 
\def \Omegam {\ifmmode \Omega_{\rm m} \else $\Omega_{\rm m}$ \fi} 
\def \Omegab {\ifmmode \Omega_{\rm b} \else $\Omega_{\rm b}$ \fi} 
\def \fbar {\ifmmode f_{\rm bar} \else $f_{\rm bar}$ \fi} 
\def \OmegaL {\ifmmode \Omega_{\rm \Lambda} \else $\Omega_{\rm \Lambda}$\fi} 
\def \Deltavir {\ifmmode \Delta_{\rm vir} \else $\Delta_{\rm vir}$ \fi}
\def \rhocrit {\ifmmode \rho_{\rm crit} \else $\rho_{\rm crit}$ \fi}

\def \ri {\ifmmode r_{\rm i} \else $r_{\rm i}$ \fi} 
\def \rf {\ifmmode r_{\rm f} \else $r_{\rm f}$ \fi}
\def \ra {\ifmmode r_{\rm a} \else $r_{\rm a}$ \fi} 
\def \Mi {\ifmmode M_{\rm i} \else $M_{\rm i}$ \fi}
\def \Ma {\ifmmode M_{\rm a} \else $M_{\rm a}$ \fi}
\def \Va {\ifmmode V_{\rm a} \else $V_{\rm a}$ \fi} 
\def \Mf {\ifmmode M_{\rm f} \else $M_{\rm f}$ \fi}
\def \Ei {\ifmmode E_{\rm i} \else $E_{\rm i}$ \fi}
\def \Ef {\ifmmode E_{\rm f} \else $E_{\rm f}$ \fi}
\def \fin {\ifmmode f_{\rm in} \else $f_{\rm in}$ \fi}
\def \fout {\ifmmode f_{\rm out} \else $f_{\rm out}$ \fi} 

\def \rs {\ifmmode r_{\rm s} \else $r_{\rm s}$ \fi} 
\def \Rvir {\ifmmode R_{\rm vir} \else $R_{\rm vir}$ \fi}
\def \Vvir {\ifmmode V_{\rm  vir} \else  $V_{\rm vir}$  \fi} 
\def \Vmax {\ifmmode V_{\rm  max} \else  $V_{\rm max}$  \fi} 
\def \Mvir {\ifmmode M_{\rm  vir} \else $M_{\rm  vir}$ \fi}
\def \Mhalo {\ifmmode M_{200} \else $M_{200}$ \fi}  

\def \eSF {\ifmmode \epsilon_{\rm SF} \else $\epsilon_{\rm SF}$ \fi}
\def \eR {\ifmmode \epsilon_{\rm R} \else $\epsilon_{\rm R}$ \fi} 
\def \Rd {\ifmmode R_{\rm d} \else $R_{\rm d}$ \fi} 
\def \Mstar {\ifmmode M_{\rm star} \else $M_{\rm star}$ \fi} 
\def \Mgas {\ifmmode M_{\rm gas} \else $M_{\rm gas}$ \fi}
\def \rhalf {\ifmmode r_{\rm 1/2} \else $r_{\rm 1/2}$ \fi} 

\title[Dark Halo Response to Galaxy Formation] {NIHAO IX: the role of
  gas inflows and outflows in driving the contraction and expansion of
  cold dark matter haloes}

\author[Dutton et al.]{Aaron  A. Dutton,$^{1}$\thanks{dutton@nyu.edu}
 Andrea V. Macci\`o$^{1,2}$,
 Avishai Dekel$^3$,
 Liang Wang$^{4}$,
\newauthor{Gregory Stinson$^2$,
 Aura Obreja$^1$,
 Arianna Di Cintio$^5$,
 Chris Brook$^6$,
}
\newauthor{Tobias Buck$^2$, Xi Kang$^4$}
\\
  $^1$New York University Abu Dhabi, PO Box 129188 Abu Dhabi, United Arab Emirates\\
  $^2$Max-Planck-Institut f\"ur Astronomie, K\"onigstuhl 17, D-69117 Heidelberg, Germany\\
  $^3$Center for Astrophysics and Planetary Science, Racah Institute of Physics, The Hebrew University, Jerusalem 91904, Israel\\
  $^4$Purple Mountain Observatory, the Partner Group of MPI f\"ur Astronomie, 2 West Beijing Road, Nanjing 210008, China\\
  $^5$DARK-Carlsberg Fellow, Dark Cosmology Centre, NBI, Juliane Maries Vej 30, DK-2100 Copenhagen, Denmark\\
  $^6$Ramon y Cajal Fellow, Departamento de F\'isica Te\'orica, Universidad Aut\'onoma de Madrid, E-28049 Cantoblanco, Madrid, Spain}

\begin{document}
             
\date{Accepted 2016 June 22. Received 2016 June 22; in original form 2016 May 17}
             
\pagerange{\pageref{firstpage}--\pageref{lastpage}}\pubyear{2016}

\maketitle           

\label{firstpage}
             

\begin{abstract}
   We use $\sim 100$ cosmological galaxy formation `zoom-in' simulations
  using the smoothed particle hydrodynamics code {\sc gasoline} to
  study the effect of baryonic processes on the mass profiles of cold
  dark matter haloes.
  The haloes in our study range from dwarf ($M_{200}\sim10^{10}\Msun$)
  to Milky Way ($M_{200}\sim10^{12}\Msun$) masses.
  Our simulations exhibit a wide range of halo responses, primarily
  varying with mass, from expansion to contraction, with up to factor
  $\sim 10$ changes in the enclosed dark matter mass at 1 per cent of
  the virial radius.
  Confirming previous studies, the halo response is correlated with
  the integrated efficiency of star formation: $\eSF\equiv
  (\Mstar/M_{200})/(\Omegab/\Omegam)$.
  In addition, we report a new correlation with the compactness of the
  stellar system: $\eR\equiv \rhalf/R_{200}$.
  We provide an analytic formula depending on $\eSF$ and $\eR$ for the
  response of cold dark matter haloes to baryonic processes.
  An observationally testable prediction is that, at fixed mass,
  larger galaxies experience more halo expansion, while the smaller
  galaxies more halo contraction.
  This diversity of dark halo response is captured by a toy model
  consisting of cycles of adiabatic inflow (causing contraction)
  and impulsive gas outflow (causing expansion).
  For net outflow, or equal inflow and outflow fractions, $f$, the
  overall effect is expansion, with more expansion with larger $f$.
  For net inflow, contraction  occurs for small $f$ (large radii),
  while expansion occurs for large $f$ (small radii), recovering the
  phenomenology seen in our simulations.
  These regularities in the galaxy formation process provide a step
  towards a fully predictive model for the structure of cold dark
  matter haloes.
\end{abstract}


\setcounter{footnote}{1}


\section{Introduction}
\label{sec:intro}

Accurately predicting the structural properties of cold dark matter
(CDM) haloes is one of the fundamental goals of modern cosmology.  A
large amount of effort has been spent studying the radial profiles of
CDM haloes as a function of halo mass, time and cosmological
parameters \citep[e.g.,][]{NFW97, Bullock01, Maccio08, Zhao09, Stadel09,
  Navarro10, Klypin11, Dutton14b, Diemer15, Ludlow16}.  If
one ignores the effects of gas dissipation and star formation, these
dissipationless simulations make robust predictions that can be used
to test, and potentially falsify, the CDM model.  One of these
predictions is that CDM haloes are roughly scale free with `cuspy'
central density profiles, $\rho(r) \propto r^{-\alpha}$, with
$\alpha\approx 1.2$.

Observations of massive elliptical galaxies, which have non-negligible
uncertainties in subtracting off the stars, favour these cuspy haloes
over cores ($\alpha\approx 0$) or contracted haloes \citep{Dutton13b,
  Dutton14a, Sonnenfeld15}.  However, observations of dark matter
dominated systems, where subtracting off the baryons is less
problematic, have yet to find convincing evidence of these
cusps. Rather, they tend to favour shallower inner density slopes
$\alpha \lta 0.5$ both in low-mass galaxies \citep[e.g.,][]{deBlok01,
  Swaters03, Goerdt06, Walker11, Oh11} and galaxy clusters
\citep[e.g.,][]{Newman13}.

These discrepancies are a potentially fatal problem for the CDM
paradigm, and has thus motivated studying other flavours of dark matter
that have similar behaviour to CDM on large scales, but differ on small
scales. Popular examples are warm dark matter \citep[WDM;][]{Pagels82}
and self-interacting dark matter \citep[SIDM;][]{Spergel00}. One can
also consider more exotic models that have a coupling between the dark
matter and dark energy \citep[e.g.,][]{Maccio15}.

Since the baryons make up $\Omegab/\Omegam\approx 15\%$
\citep{Planck14} of the mass in the Universe, to first order it might
seem reasonable to assume that baryons only have a small effect on the
structure of dark matter haloes.  However, one needs to consider that
baryons and dark matter are not equally mixed.  Baryons can dissipate
their energy, leading to higher baryon to dark matter ratios in the
centres of galaxies.  It is not uncommon for the baryons to dominate
the mass budget inside galaxy half-light radii \citep[e.g., Fig.2
  of][]{Courteau15}.

Baryonic processes are known that can cause both expansion and
contraction of dark matter haloes at small radii.  If the accretion of
baryons on to the central galaxy is smooth and slow then dark matter
haloes should contract adiabatically \citep{Blumenthal86, Gnedin04}.
This process can increase the density of dark matter haloes by an
order of magnitude.  In contrast, other processes can cause the dark
matter halo to expand: rapid mass-loss and/or time variability of the
potential due to feedback from stars \citep{Navarro96, Read05,
  Mashchenko06, Pontzen12, Maccio12, El-Zant16} or Active Galactic
Nuclei \citep[AGN;][]{Martizzi12, Martizzi13}; mergers and stripping
of smaller subhaloes that were puffed up by stellar feedback
\citep{Dekel03}; and transfer of energy/angular momentum from baryons
to the dark matter via dynamical friction due to minor mergers
\citep{El-Zant01, Jardel09, Johansson09, Cole11}, or galactic bars
\citep{Weinberg02, Sellwood08}.

Supernova feedback that can alter galactic potentials plays other
important roles in the formation of galaxies.  It is the principle
explanation for why galaxy formation is so inefficient in haloes of
Milky Way (MW) virial mass ($M_{200}\sim 10^{12}\Msun$) and below
\citep[e.g.,][]{Dekel86, Hopkins14, Wang15}. Stellar feedback also
ejects low angular momentum material, enabling the formation of
exponential disc-dominated galaxies \citep[e.g.,][]{Maller02,
  Dutton09, Brook11, Brooks16}.  Likewise, AGN
feedback is the principle explanation for why galaxy formation is so
inefficient in haloes more massive than the MW
\citep[e.g.,][]{Croton06, Somerville08}.

At the mass scale of the MW early cosmological hydrodynamical
simulations  almost invariably found that dark matter haloes contract
in response to galaxy formation \citep[e.g.,][]{Gustafsson06,
  Romano-Diaz08, Abadi10,  Tissera10, Duffy10}.  However,
these simulations generally find that the contraction is weaker than
predicted by the \citet{Blumenthal86} and \citet{Gnedin04} models. A
potentially serious problem with these simulations is they over
predict the stellar masses of galaxies, and thus it is not clear how
applicable they are to galaxies in the real Universe.  More recent
simulations, with stronger feedback and hence more realistic stellar
to halo mass rations, report weak contraction or no change in the dark
matter mass profiles on MW mass scales
\citep{Marinacci14,Schaller15}.

On the scale of dwarf galaxies ($M_{200}\sim 10^{10-11}\Msun$), a
number of different groups have presented  fully cosmological
hydrodynamical simulations in which the dark matter halo expands
\citep{Mashchenko08, Governato12, Trujillo-Gomez15, Chan15, Tollet16}.
In these simulations stellar feedback was found to drive rapid
(sub-dynamical) time variability of the potential. Dark matter
particles orbiting in such a potential gain energy, resulting in halo
expansion \citep{Pontzen12}.  The use of different hydrodynamical
codes and sub-grid models suggests that halo expansion is not a
numerical artefact, or due to a specific modelling of the subgrid
physics.

These apparently conflicting results on dwarf and MW mass
scales can be reconciled if the halo response depends, on average, on
the halo mass, $M_{200}$, galaxy stellar mass, $\Mstar$, or   the
integrated efficiency of star formation,
\begin{equation}
  \eSF\equiv (\Mstar/M_{200})/(\Omegab/\Omegam).
\end{equation}
\citet{DiCintio14a} found that the inner dark matter density slope,
$\alpha$, was more strongly correlated with $\eSF$, than galaxy or
halo mass.  At very low efficiencies, there is no change with
$\alpha\approx 1.2$ as would be expected.  As the star formation
efficiency increases, $\alpha$ decreases, reaching $\alpha\approx 0$
at $\eSF \sim 3\%$. For even higher efficiencies $\alpha$ increases,
reaching the CDM value of $\alpha \approx 1.2$ at an efficiency of $\sim
20\%$, and steeper values as high as $\alpha\approx 2$ for
efficiencies approaching unity.

\begin{table*}
\begin{center}
  \caption{Summary of the main differences MUGS, MaGICC and NIHAO simulations, see text for discussion.}
\label{tab:param}
  \begin{tabular}{cccccccccc}
  \hline
  \hline
name & {\sc gasoline} & cosmology & $n_{\rm th}$ & $c_*$ & $\Delta t$ & $E_{\rm SN}$    & $\epsilon_{\rm ESF}$ & IMF\\ 
&      &           & (cm$^{-3}$)  &      & (Myr)      & ($10^{51}$erg) &                     &     \\
\hline  
MUGS  & \citet{Wadsley04}  & {\it WMAP}3  & $\phantom1$0.1      & 0.05 & 1.0 & 0.4   &    0.0$\phantom3$ &  \citet{Kroupa93}   & \\
MaGICC & \citet{Wadsley04} & {\it WMAP}3  & $\phantom1$9.3      & 0.1$\phantom5$ & 0.8 & 1.0   &    0.1$\phantom3$ &  \citet{Chabrier03}   & \\
NIHAO & \citet{Keller14} & Planck & 10.6     & 0.1$\phantom5$ & 0.8 & 1.0   &    0.13         &  \citet{Chabrier03}   & \\
 \hline
\end{tabular}
\label{tab:parameters}
\end{center}
\end{table*}

\citet{DiCintio14b} present a parametrization of the dark matter
density profiles in the MaGICC simulations \citep{Stinson13}, using a
five parameter function. As with the inner slope, the other parameters
were found to depend on $\eSF$.  The goal of this paper is slightly
different.  Rather than parameterizing the dark matter halo in the
hydro simulations, we wish to describe how the dark matter changes
[with respect to dark matter-only (DMO) simulations] as a function of the
baryonic mass distribution in a galaxy. We also use an order of
magnitude more simulations than MaGICC, so we can explore variation of
halo response to multiple parameters.  Ultimately we want to
understand the physical mechanisms that cause contraction and
expansion, and to model them.  If a physical regularity can be found,
it will enable a more general prediction of CDM density
profiles in semi-analytic galaxy formation models and when mass
modelling galaxy kinematic observations. 

This paper is organized as follows. The galaxy formation simulations
we use are presented in \S\ref{sec:simulation}. Changes to the dark
matter profiles (between hydrodynamical and dissipationless simulations)
are given in \S\ref{sec:profiles}, including an
analytic fitting formula that describes these results in
\S\ref{sec:formula}. \S\ref{sec:physicalmodel} presents results for
the halo response of our simulations in the context of the adiabatic
contraction formalism, and includes a toy model for halo response
involving cycles of inflows and outflows in \S\ref{sec:toymodel}.  We
discuss our results in \S\ref{sec:discussion}, and give a summary in
\S\ref{sec:summary}.


\section{Simulations} \label{sec:simulation}

The galaxies we analyse here come from three projects: Numerical
Investigation of a Hundred Astrophysical Objects
\citep[NIHAO;][]{Wang15}, Making Galaxies in a Cosmological Context
\citep[MaGICC;][]{Stinson13}, and McMaster Unbiased Galaxy Simulations
\citep[MUGS;][]{Stinson10}.   As described below, NIHAO and MaGICC use
a similar star formation and feedback model, and form the `correct'
amount of stars. MUGS galaxies are too efficient at forming stars, but
we use them as they provide a limiting case for when star formation is
very efficient, and they provide a link to similar early simulation
works. The main parameters that differ between the simulations are given
in Table~\ref{tab:param}.

All three sets of simulations cover the halo mass range $\sim 10^{10}
- 10^{12} \Msun$ at redshift $z=0$.  NIHAO consists of 78 galaxies,
MaGICC 11, and MUGS 5.  MaGICC and MUGS use the same initial
conditions of MW mass haloes, and re-scale them to simulate
lower mass galaxies. NIHAO selects haloes to have roughly uniform
sampling in log halo mass, and independent of formation history at a
given halo mass.

As discussed in previous papers, MaGICC and NIHAO form some of the
most realistic simulated galaxies to date, being consistent with a
wide range of galaxy properties both in the local and distant
universe: evolution of stellar to halo masses and star formation rates
\citep{Stinson13, Wang15}; disc sizes \citep{Brook12}, gas phase
metallicities \citep{Obreja14}; the shape of the MW's dark
matter halo \citep{Butsky15}; cold and hot gas content
\citep{Stinson15, Gutcke16, Wang16}; {\sc Hi} sizes and the baryonic
Tully-Fisher relation \citep{Brook16}; stellar disk kinematics
\citep{Obreja16}; shallow inner dark matter density slopes of dwarf
galaxies \citep{Tollet16};  and resolve the `too-big-to-fail'
problem of nearby field dwarf galaxies \citep{Dutton16}.  They thus
provide plausible, if not unique, templates for how baryonic processes
effect the structure of CDM haloes.

\subsection{Cosmology}
MUGS and MaGICC uses the {\it WMAP}3 cosmology \citep{Spergel07}  with
Hubble parameter $H_0$= 73.0 \kms Mpc$^{-1}$, matter density
$\Omegam=0.24$, dark energy density $\Omega_{\Lambda}=1-\Omegam=0.76$,
baryon density $\Omegab=0.04$ and power spectrum normalization
$\sigma_8 = 0.76$, and power spectrum slope $n=1$, while NIHAO
uses the Planck cosmology \citep{Planck14}: Hubble parameter $H_0$=
67.1 \kms Mpc$^{-1}$, matter density $\Omegam=0.3175$, dark energy
density $\Omega_{\Lambda}=1-\Omegam=0.6825$, baryon density
$\Omegab=0.0490$, power spectrum normalization $\sigma_8 = 0.8344$,
power spectrum slope $n=0.9624$.

\subsection{Hydrodynamics}
All three simulation sets use the {\sc gasoline} \citep{Wadsley04}
smoothed particle hydrodynamics (SPH) code. MUGS and MaGICC use the
original version, while NIHAO uses the version of {\sc gasoline} from
\citet{Keller14} that reduces the formation of blobs and improves
mixing. Both versions of {\sc {gasoline}} use the same treatment for
cooling via hydrogen, helium, and various metal-lines in a uniform
ultraviolet ionizing background as described in \citet{Shen10}.

\subsection{Star formation}
The simulations use a common recipe for star formation following
\citet{Stinson06}, but with three different sets of parameters.  Stars
form from cool ($T < 15 000K$) dense gas ($n > n_{\rm th}$).  A subset
of the particles that pass these criteria are randomly selected to
form stars based on the commonly used star formation equation,
$\Delta{m_*}/\Delta t = c_* m_{\rm gas}/t_{\rm dyn}$. Here, $\Delta
m_*$ is the mass of the star particle formed, $\Delta t$ is the
time step between star formation events, and $c_*$ is the efficiency of
star formation, i.e., the fraction of gas that will be converted into
stars during a dynamical time.

In MUGS the threshold density is $n_{\rm th}=0.1 \,{\rm cm^{-3}}$,
$c_*=0.05$, and $\Delta t=1$ Myr.  In MaGICC and NIHAO $\Delta
t=0.8$Myr, $c_*=0.1$ and  $n_{\rm th}$ is set to the maximum density
at which gravitational instabilities can be resolved $n_{\rm th}=
n_{\rm sph} m_{\rm gas} / \epsilon_{\rm gas}^3$.  Here $n_{\rm sph}$
is the number of particles in the SPH smoothing kernel, $m_{\rm gas}$
is the gas particle mass, and $\epsilon_{\rm gas}$ is the force
softening of the gas particles. The threshold density is 9.3 cm$^{-3}$
in MaGICC and 10.6 cm$^{-3}$ in NIHAO.

\subsection{Feedback}

The thermal stellar feedback comprises two epochs. The first epoch,
`early stellar feedback' (ESF), happens before any supernovae
explode. It represents stellar winds and photoionization from the
bright young stars.  The  ESF consists of a fraction $\epsilon_{\rm
  ESF}$ of the total stellar flux being ejected from stars into
surrounding gas ($2 \times 10^{50}$ erg of thermal energy per $\Msun$
of the entire stellar population).  Radiative cooling is left {\it on}
for the ESF.  MUGS does not include ESF (i.e., $\epsilon_{\rm
  ESF}=0$). MaGICC adopts $\epsilon_{\rm ESF}=0.10$, which has been
tuned to match the evolution of the stellar mass versus halo mass
relation \citep{Behroozi13} for a $z=0$ MW mass galaxy
\citep{Stinson13}.  NIHAO adopts $\epsilon_{\rm ESF}=0.13$. The slight
change is due to the increased mixing, and subsequent higher star
formation efficiency, in the upgraded version of {\sc gasoline} used
in NIHAO. 

The second epoch starts 4 Myr after the star forms, when the first
supernovae start exploding. Only supernova energy is considered as
feedback in this second epoch.  Supernova feedback is implemented
using the blastwave formalism described in \citet{Stinson06}.  Since
the gas receiving the energy is dense, it would quickly be radiated
away due to its efficient cooling. For this reason, cooling is delayed
for particles inside the blast region for $\sim 30$ Myr.  Stars $8
\Msun < m_{\rm star} < 40 \Msun$ eject both energy, $E_{\rm SN}$, and
metals into the interstellar medium gas surrounding the region where
they formed. MUGS adopts $E_{\rm SN}=4\times 10^{50}$ erg per SN, and
a \citet{Kroupa93} initial mass function (IMF), while MaGICC and NIHAO
adopt $E_{\rm SN}=1\times 10^{51}$ erg, and a \citet{Chabrier03}
IMF. The different IMF and energy per supernova result in a factor of
$\approx 5$ more energy released per solar mass of stars formed in the
MaGICC and NIHAO simulations.  This feedback is sufficiently strong to
correct the over cooling problem of the MUGS simulations.

\begin{figure}
\centerline{
  \psfig{figure=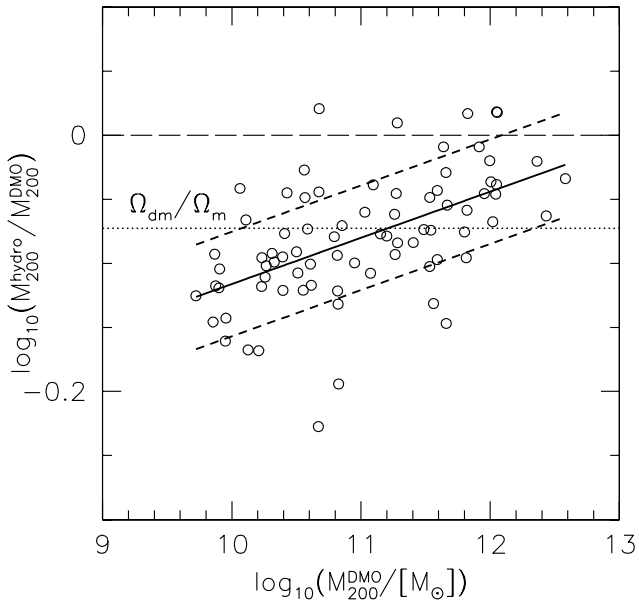,width=0.48\textwidth} 
}
\caption{Comparison between halo masses, $M_{200}$, in hydro and DMO
  NIHAO simulations. The long-dashed line represents equality, showing
  that most hydro simulations have lost mass relative to the DMO
  simulations. In low-mass haloes, the mass-loss is greater than simply
  failing to accrete or retain the cosmic baryon fraction of mass
  (dotted line). }
\label{fig:mvir}
\end{figure}

\subsection{Halo identification}

The main haloes in the simulations were identified using the
MPI+OpenMP hybrid halo finder
\texttt{AHF}\footnote{http://popia.ft.uam.es/AMIGA}
\citep{Gill04,Knollmann09}. \texttt{AHF} locates local over-densities
in an adaptively smoothed density field as prospective halo
centres. The virial masses of the haloes are defined as the masses
within a sphere containing $\Delta=200$ times the cosmic critical
matter density, $\rho_{\rm crit}=3H(z)^2 / 8 \pi G$.

Each hydrodynamic simulation has a corresponding DMO
simulation. We use the superscripts `hydro' and `DMO' to refer to
parameters from the hydrodynamical, and DMO simulations,
respectively. For example, the halo masses are denoted $M_{200}^{\rm
  hydro}$ and $M_{200}^{\rm DMO}$. In general the halo masses for a
given initial condition are not identical between the hydro and DMO
simulations (Fig.~\ref{fig:mvir}).  In order for a meaningful
comparison between the dark matter haloes profiles we have removed
halo pairs with significantly different masses, which for example,
occurs when one of the haloes is unrelaxed due to a recent major
merger, or when a major merger has not occurred yet in one of the
simulations.  Loss of gas due to feedback results in a systematic
trend with halo mass:
\begin{equation}
  \log_{10}\left( \frac{M_{200}^{\rm hydro}}{M_{200}^{\rm DMO}}\right) = -0.0447 +0.0369\log_{10}\left(\frac{M_{200}^{\rm DMO}}{10^{12}\Msun}\right),
\end{equation}
with a scatter of 0.039 dex.
Due to these differences, when we
calculate quantities at a fraction of the virial radius we pick that
of the DMO simulation to ensure that we are using the same physical
radius for each simulation pair.   Note that the low-mass haloes have
lost more mass than simply removing the baryons according to the
cosmic baryon fraction (dotted line in Fig.~\ref{fig:mvir}),
indicating a slowing down of the mass accretion histories of the hydro
haloes. Similar results have been found in the APOSTLE simulations of
Local Group analogues by \citet{Sawala16}.

\begin{figure*}
  \centerline{
     \psfig{figure=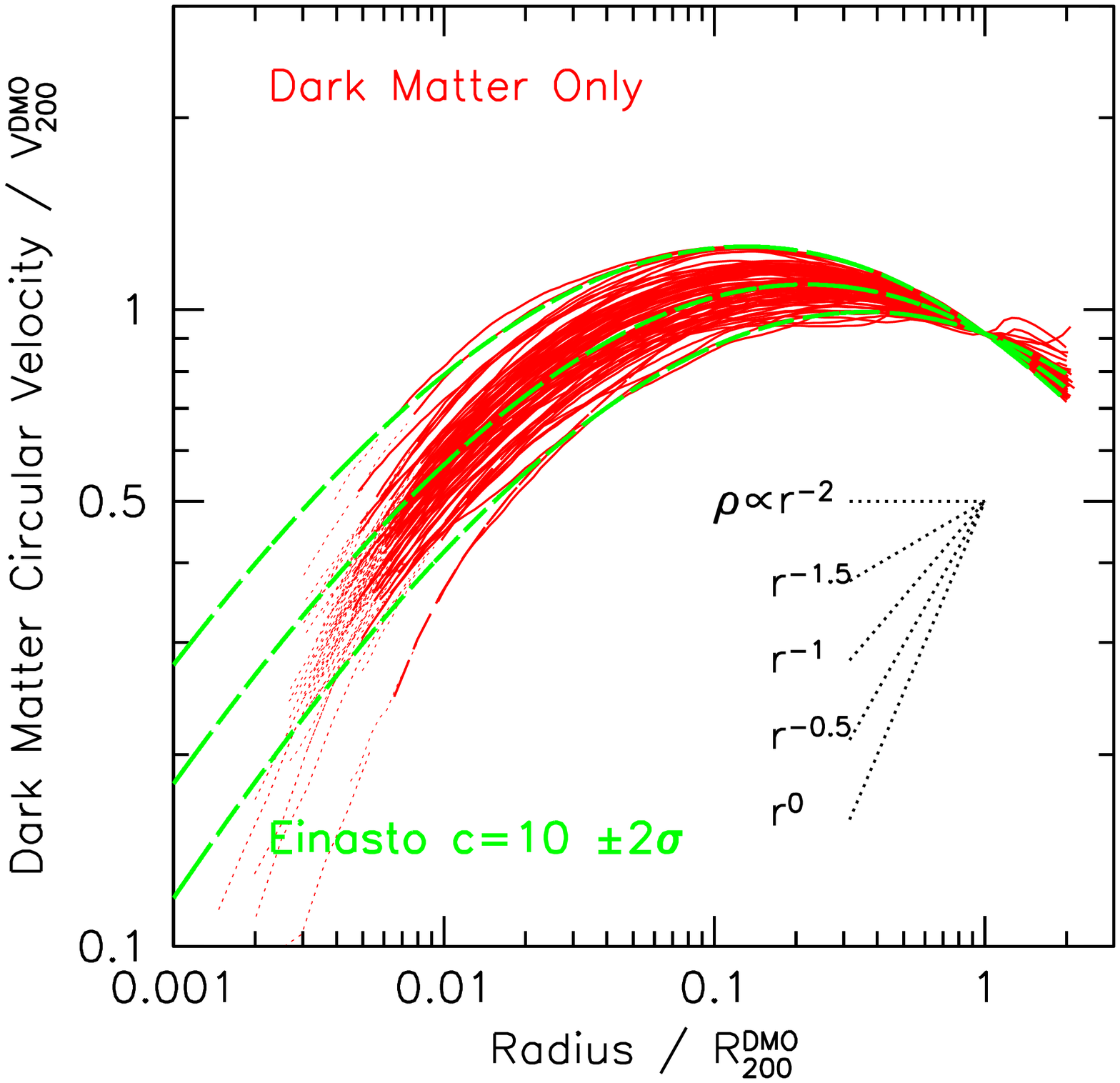,width=0.48\textwidth} 
     \psfig{figure=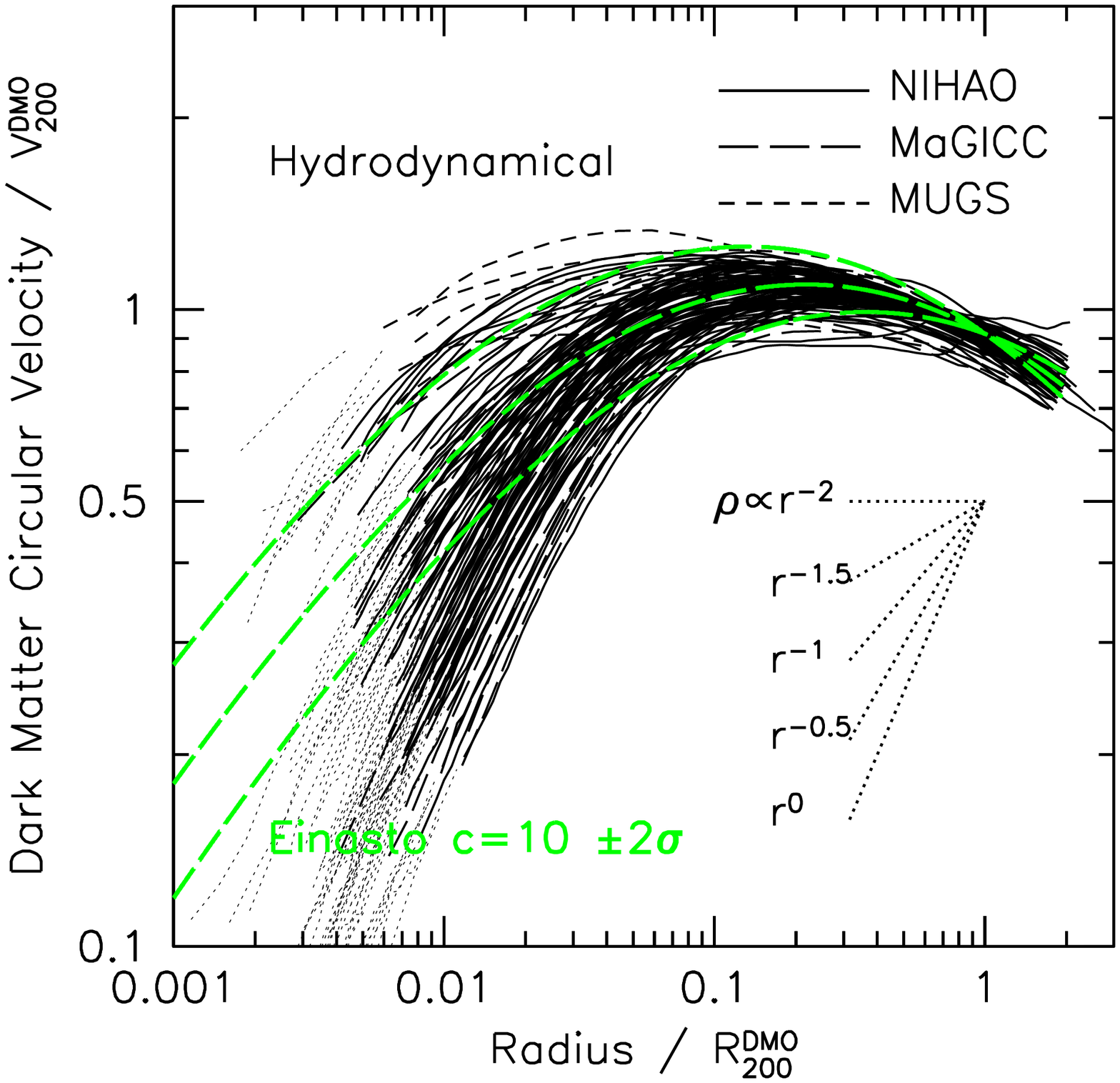,width=0.48\textwidth}
}
\caption{Dark matter circular velocity curves normalized to the virial
  velocity and virial radius of the DMO simulations. For the DMO
  simulations we have scaled the velocities by $\sqrt{1-\fbar}$ to
  make them directly comparable to the hydrodynamical simulations.
  Solid and dashed lines show the profiles down to where the velocity
  profile has converged to 10\%, while the dotted lines continue the
  profile to the dark matter softening length. Hydro simulations
  (right-hand panel) show a much larger spread in rotation curve
  shapes than the DMO simulations (left-hand panel).  For reference,
  Einasto profiles with concentration $c=6,10,16.6$ (the $2\sigma$
  range) and $\alpha=0.18$ are shown (long-dashed green). Circular
  velocity slopes for various power-law density profiles (black
  dotted) show that the hydro simulations can have both steeper and
  shallower density profiles than those found in DMO simulations. }
\label{fig:vv}
\end{figure*}


\section{Dark Matter Profiles}
\label{sec:profiles}

\subsection{Mass profiles}
The dark matter circular velocity profiles, $V_{\rm circ}=\sqrt{G
  M(<r)/r}$, for our cosmological simulations are shown in
Fig.~\ref{fig:vv}. The radii and velocities have been scaled to the
virial radius, $R_{200}^{\rm DMO}$, and virial velocity, $V_{200}^{\rm
  DMO}$, of the DMO simulations. In addition, we have scaled the
velocities of the DMO simulations by $\sqrt{1-\fbar}$, where $\fbar
\equiv \Omegab/\Omegam$, to approximate the contribution of baryons to
the DMO velocity profile.  The left-hand panel shows results for the
dissipationless simulations, while the right-hand panel shows the
results for the hydro simulations.  In both panels, the solid lines
show simulations from NIHAO, while the dashed lines show simulations
from MaGICC and MUGS.  These lines are only shown for radii where the
circular velocity profile is converged to 10\% according to the
convergence criteria of \citet{Power03}, modified by
\citet{Schaller15}.  Dotted lines show the velocity profiles down to
the softening length of the dark matter particles, which is typically
two to three times smaller than the convergence radius.

As expected, the DMO simulations are well described by the Einasto
profile \citep{Einasto65},
\begin{equation}
\ln[\rho(r)/\rho_{-2}] = -(2/\alpha) [(r/r_{-2})^{\alpha}-1],
\end{equation}
with $\alpha=0.18$ and concentrations $c\equiv
R_{200}/r_{-2}=6,10,16.6$ corresponding to the $2\sigma$ range found
in large volume $N$-body simulations \citep{Dutton14b}.  However, at the
smallest resolved radii many of the hydro simulations have dark matter
density profiles with inner slopes shallower than the nominal CDM
value of $r^{-1}$.  There is also clearly a much wider range in dark matter
profiles in the hydro simulations. For example, at 1\% of the virial
radius, the DMO simulations show an absolute variation in
dark matter circular velocity of just a factor of 2, whereas the hydro
simulations show an order of magnitude spread. 

\begin{figure*}
  \centerline{
     \psfig{figure=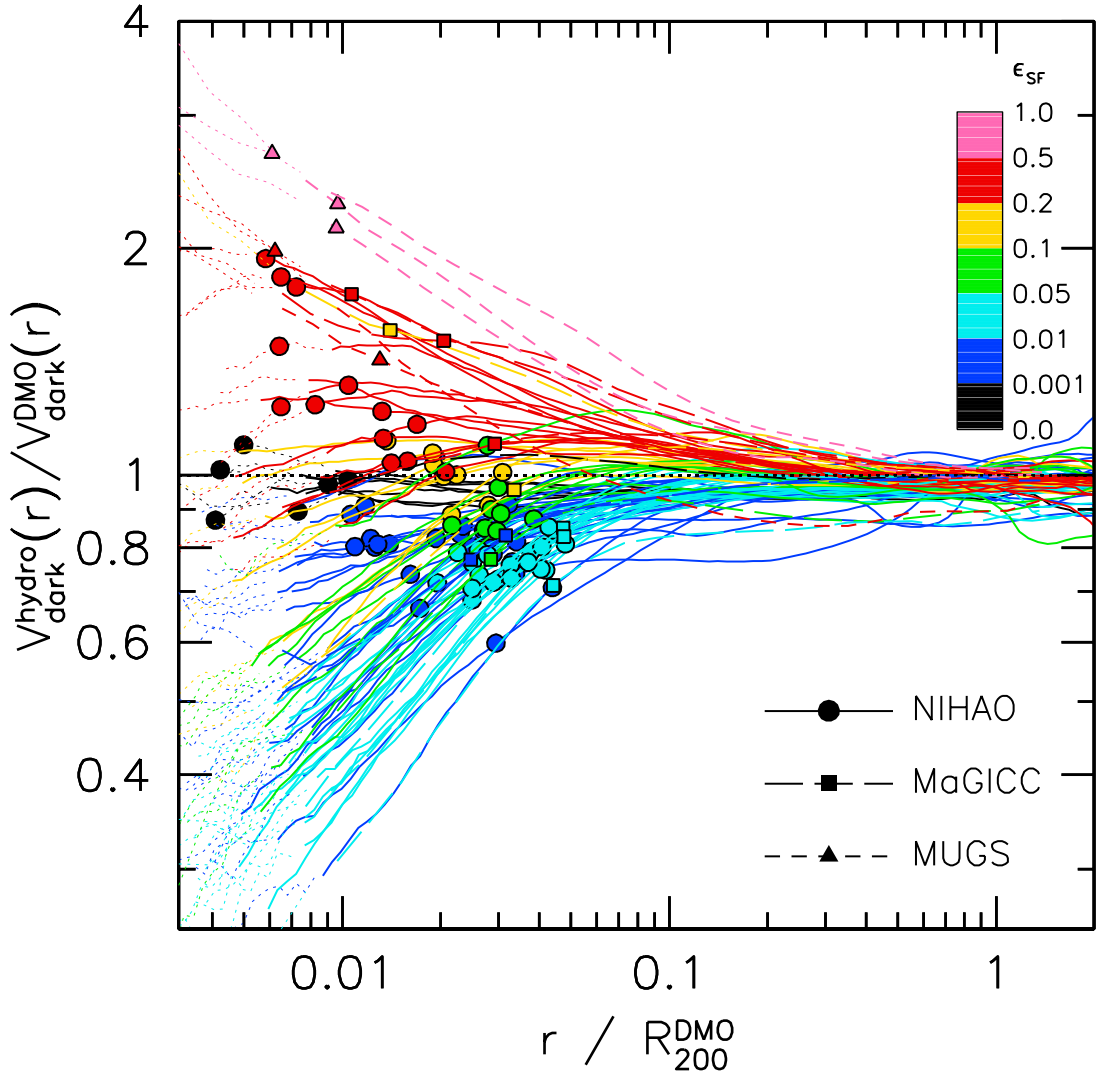,width=0.85\textwidth} } 
  \caption{Ratio between dark matter circular velocity curves from the
    hydro $V_{\rm dark}^{\rm hydro}(r)$ and dissipationless $V_{\rm
      dark}^{\rm DMO}$ simulations.  Points show the location of
    stellar half-mass radii, with different symbols for the NIHAO
    (circles), MaGICC (squares), and MUGS (triangles) simulations.
    Colours corresponds to integrated star formation efficiency,
    $\Mstar/M_{\rm 200}/\fbar$, as indicated.}
\label{fig:vv2}
\end{figure*}

\subsection{Halo response to galaxy formation}
\label{sec:haloresponse}
The change in dark matter profiles is  shown more clearly in
Fig.~\ref{fig:vv2}. The $y$-axis shows the ratio between the dark matter
circular velocity in the hydro simulations, $V_{\rm dark}^{\rm
  hydro}(r)$, to that in the DMO, $V_{\rm dark}^{\rm
  DMO}(r)$, simulations. The DMO velocities have been
scaled by a factor of $\sqrt{1-\fbar}$ so that a ratio of 1 implies no
change in the dark matter mass profile. The mass ratio is thus
$[V_{\rm dark}^{\rm hydro}(r)/V_{\rm dark}^{\rm DMO}(r)]^2$. For a
power-law density profile $\rho\propto r^{-1}$ the enclosed mass goes
like $M(r)\propto r^{2}$, and thus the velocity ratio roughly
corresponds to the radial expansion (or contraction factor). For
example, a velocity ratio of 0.5 means shells of dark matter have
expanded a factor of $\sim 2$.

The solid lines show simulations from NIHAO, while the dashed lines
show simulations from MaGICC and MUGS.  The points show the location
of the 3D half-mass radius of the stars: NIHAO (circles), MaGICC
(squares), and MUGS (triangles). The MaGICC and NIHAO simulations span
a similar range of halo response, while the MUGS simulations are among
the most contracted.  Amongst other things, this shows that the role
of the hydrodynamics (which has been improved in NIHAO versus MaGICC)
plays a subdominant role to the strength of the feedback.

Changes are larger at smaller radii, both when the haloes expand and
contract.  At radii greater than $10\%$ of the virial radius, there is
little change in the dark matter profile. This is to be expected,
since the half-mass radii of the stars are typically between 0.5 and
5 per cent of the virial radius. At these scales, where baryons condense and
stars form, there is a wide range of halo responses. At 1\% of the
virial radius the halo circular velocity increases and decreases by up
to factor of $\sim 2.5$, corresponding to a change in enclosed mass by
a factor of $\sim 6$. Clearly, baryons can have a major impact on the
structure of the dark matter halo.

\begin{figure*}
  \centerline{
     \psfig{figure=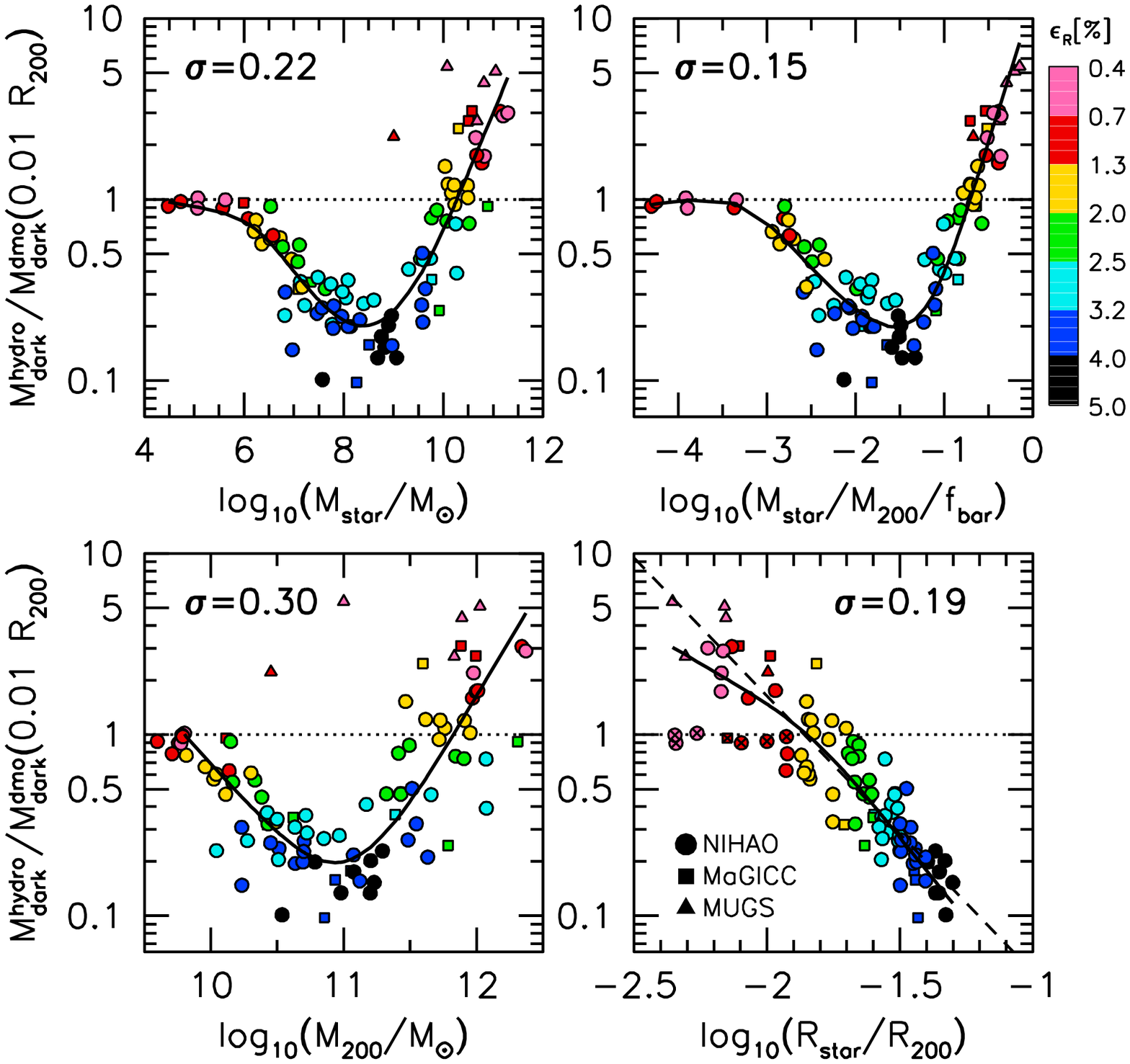,width=0.85\textwidth} } 
       \caption{Halo response at 1 per cent of the virial radius
         (expressed in terms of the dark matter mass ratio) versus
         halo mass $M_{200}$ (bottom left), stellar mass $\Mstar$ (top
         left), integrated star formation efficiency
         $\Mstar/M_{200}/\fbar$ (top right), and galaxy contraction
         factor $\eR=r_{1/2}/R_{200}$ (bottom right). Points are colour
         coded by $\eR$.  At fixed halo mass, stellar mass, and star
         formation efficiency we see a  trend for smaller galaxies to
         have more contraction / less expansion.  Point types
         correspond to the different simulation sets: NIHAO (circles),
         MaGICC (squares), MUGS (triangles).}
\label{fig:vv3}
\end{figure*}

This plot also highlights the importance of spatial resolution. In order to
see significant expansion or contraction, simulations need to be able
to resolve scales significantly smaller than $\sim 5\%$ of the virial
radius. Our simulations resolve $\sim 1\%$ of the virial radius across
three decades in halo mass. However,  large volume simulations such as
EAGLE \citep{Schaye15} and ILLUSTRIS \citep{Vogelsberger14} only have
sufficient resolution for haloes of MW mass and above.

Using the MaGICC and MUGS simulations, \citet{DiCintio14a} found that the
logarithmic slope of the dark matter density profile measured between
1 and 2\% of the virial radius was correlated with the efficiency of
star formation, $\eSF=(\Mstar/M_{200})/(\Omegab/\Omegam)$.  Thus, we colour
code the lines in Fig.~\ref{fig:vv2} according to $\eSF$. 
We identify four qualitative behaviours in the halo response:
\begin{itemize}
\item (A) no change in the dark matter profile ($\eSF \lta 10^{-3}$).
\item (B) expansion at all radii, with more expansion at smaller radii ($10^{-3} \lta \eSF \lta 0.1$) and a maximum expansion for $\eSF\sim 0.03$.
\item (C) contraction at large radii with a maximum near $0.05
  R_{200}$, together with expansion at small radii ($0.1 \lta \eSF \lta
  0.2$).
\item (D) contraction at all radii with more contraction at smaller radii ($\eSF \gta 0.2$).
\end{itemize}

Observational estimates from halo abundance matching and weak
gravitational lensing find the maximum $\eSF \sim 0.25$ at a halo mass
of $\sim 10^{12}\Msun$ \citep[e.g.,][]{Conroy09, Moster10,
  Leauthaud12, Hudson15}. Thus case (D) is likely to occur at the peak
of star formation efficiency.  Since there is a scatter of $\sim 0.2$
dex in stellar mass at fixed halo mass \citep{More11,  Behroozi13,
  Reddick13}, galaxies with $\eSF > 0.2$ are also possible in haloes
above and below $M_{200}\sim 10^{12}\Msun$ that scatter up in
efficiency. The MW is an example of a galaxy with a high star
formation efficiency of $\eSF\simeq 0.30$.  Galaxies with very
efficient star formation $\eSF \gta 0.5$ are also possible with
$2\sigma$ deviations in $M_{200}\sim 10^{12}\Msun$ haloes, so that
even the `overcooled' simulations from MUGS may provide a template for
a subset of the galaxy population.  At the low mass end of our
simulations, haloes of mass $M_{200}=10^{10}\Msun$ are expected to
host galaxies of stellar mass $\sim 10^6$ so that $\eSF \sim
10^{-3}$. The scatter in $\eSF$ is quite possibly very large in these
low mass haloes \citep{Wang15, Garrison-Kimmel16} and thus we expect
some low-mass haloes to show no change (case A), while others
expansion (case B).

\begin{figure*}
  \centerline{
    \psfig{figure=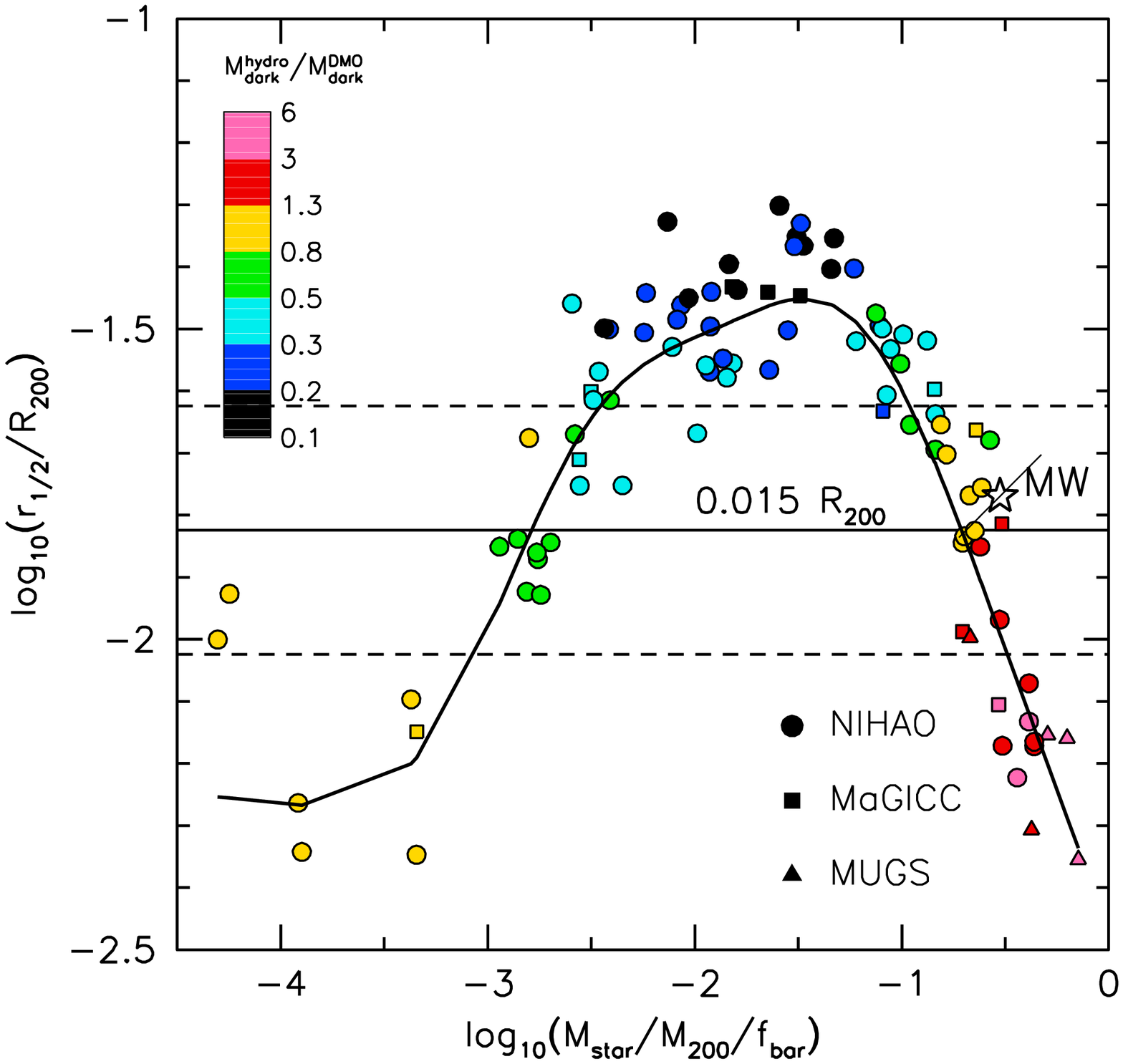,width=0.48\textwidth}
  \psfig{figure=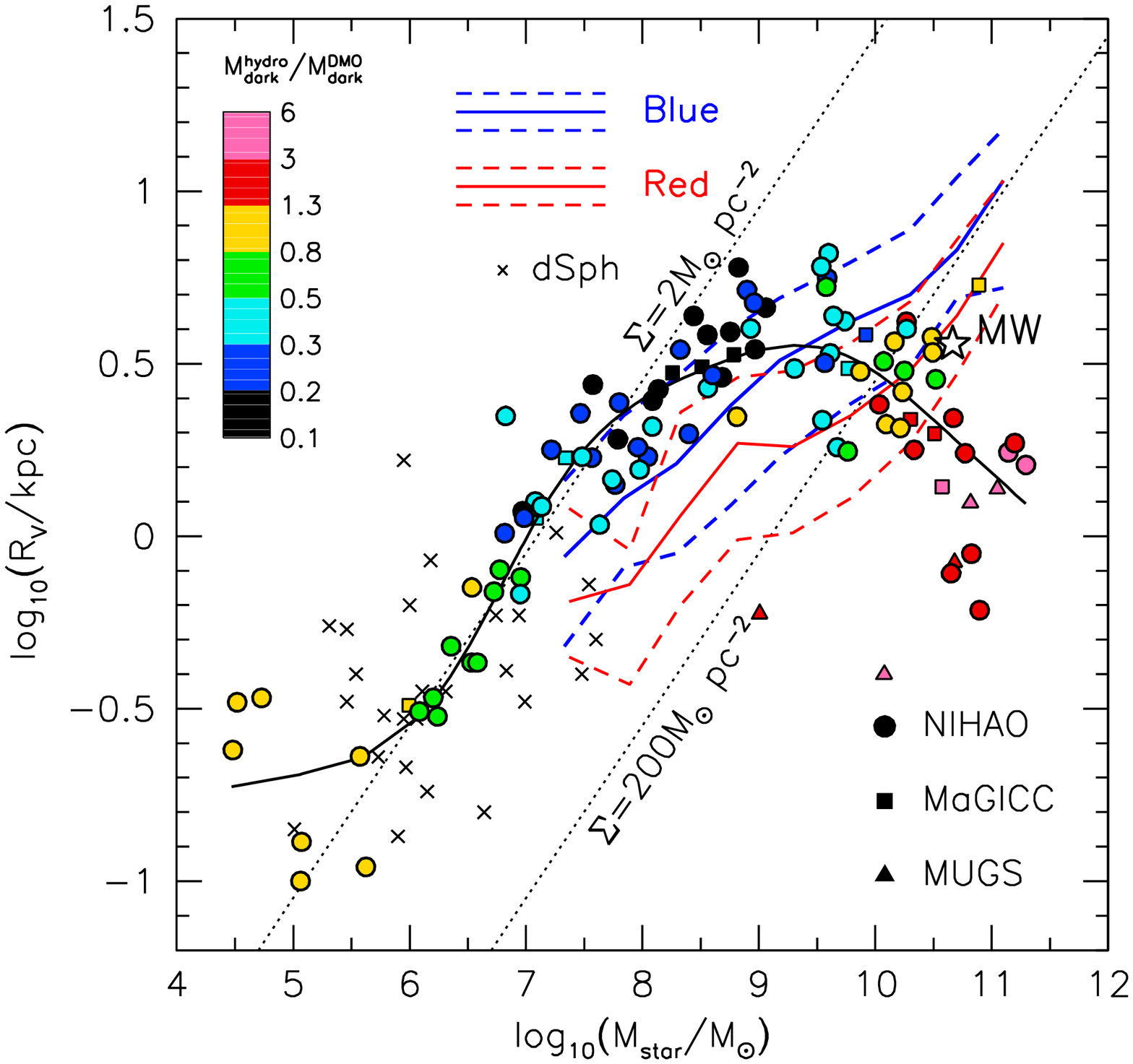,width=0.48\textwidth} }
  \caption{Halo response in the size versus stellar mass plane, using
    physical units (right) and relative to the halo size and mass
    (left).  The left-hand panel uses the 3D stellar half-mass radii,
    $\rhalf$, while the right panel uses the more directly observable
    2D stellar half-light radii in the $V$ band, $R_{V}$.  Coloured
    points show simulated galaxies from NIHAO (circles), MaGICC
    (squares), and MUGS (triangles). Points are colour coded
    by the halo response, defined to be the dark matter mass ratio
    between hydro and DMO simulations at 1\% of the virial radius. In
    the left-hand panel, the solid and dashed lines show
    $\rhalf=0.015R_{200} \pm 0.2$ dex derived by \citet{Kravtsov13}
    using halo abundance matching. In the right-hand panel, we include
    observations of 2D half-light radii from \citet{Baldry12}. Red and
    blue galaxies are shown with correspondingly coloured lines (median
    together with 16th and 84th percentiles), dwarf spheroidal (dSph)
    galaxies are shown with black crosses. The star shows the MW.
    For reference, the black dotted line show lines of
    constant effective surface density of $\Sigma=2\,\Msun\pc^{-2}$
    and  $\Sigma=200\,\Msun\pc^{-2}$.} 
\label{fig:rm}
\end{figure*}

\subsection{Correlations between halo response and galaxy parameters}
\label{sec:formula}
To show the trend with $\eSF$ more explicitly, Fig.~\ref{fig:vv3}
shows the dark matter mass ratio ($\equiv$ dark matter velocity ratio
squared) measured at 1 per cent of the virial radius versus $\eSF$ as
well as halo mass, $M_{200}$, stellar mass, $\Mstar$, and the
compactness parameter, $\eR=r_{1/2}/R_{200}$, the ratio between the 3D
stellar half-mass size and halo virial radius.  For each relation the
solid line shows a spline interpolated median, and the overall
standard deviation about each relation is given.  Of these four
fundamental parameters star formation efficiency is the best predictor
of halo response ($\sigma=0.15$), followed by compactness
($\sigma=0.19$), stellar mass ($\sigma=0.21$) and halo mass
($\sigma=0.29$).

Even though there are clear trends with stellar and halo mass in the
NIHAO (circles) and MaGICC (squares) simulations, the MUGS
simulations (triangles) show that stellar or halo mass by
themselves are not the driving parameters for halo response.  The two
lowest mass MUGS simulations have halo  masses of $M_{200}\approx
10^{10.5}$ \& $10^{11} \Msun$ and strongly contracted haloes. At these
mass scales the NIHAO simulations have the opposite halo response,
i.e., strong expansion.  Likewise the stellar masses in the MUGS simulations
are $\Mstar \approx 10^9$ \& $10^{10}\Msun$. At these mass scales, the
NIHAO simulations have a factor of 5 lower dark halo masses within 1\%
of the virial radius. By contrast, when we look at the star formation
efficiency (upper-right panel), the MUGS and NIHAO simulations follow
the same relation.

From a theoretical point of view, one would expect that the halo
response would depend not just on global properties, such as
$\Mstar/\Mhalo$, but on local ones, such as the distribution of the
baryons. In the adiabatic contraction formalism \citep{Blumenthal86}
and its variants \citep{Gnedin04, Abadi10, Dutton15}, at a fixed halo
and stellar mass, more compact baryon distributions result in more
contraction.  For the expansion case the stars are collisionless, so
they are expected to expand with the dark matter. Thus an expanded
stellar distribution is a signature of dark halo expansion. 

The lower-right panel of Fig.~\ref{fig:vv3} indeed shows a strong
correlation between the compactness of the stellar distribution,
$\eR$ and the halo response, with more extended stellar
distributions having stronger expansion.  The points in all four
panels are colour coded by $\eR$.  This shows that at fixed halo mass,
stellar mass, and star formation efficiency, smaller galaxies have
more contraction and larger galaxies have more expansion.  This
correlation fails to hold for the no-change case at low stellar
masses/star formation efficiencies ($\eSF \lta 10^{-3}$).  Excluding
these (marked with an x) and fitting a straight line gives
\begin{equation}
  y = -0.28(\pm0.11) -1.52(\pm0.42) (x+1.68)
\end{equation}
where $y=\log_{10}(M_{\rm dark}^{\rm hydro}/M_{\rm dark}^{\rm DMO})$, and
$x=\log_{10}(\rhalf/R_{200})$. The scatter about this relation
is $\sigma=0.17$.
Thus including the distribution of the stars should
improve the predictive power of a  halo response model based on global
galaxy and halo parameters.

\begin{figure*}
  \centerline{
    \psfig{figure=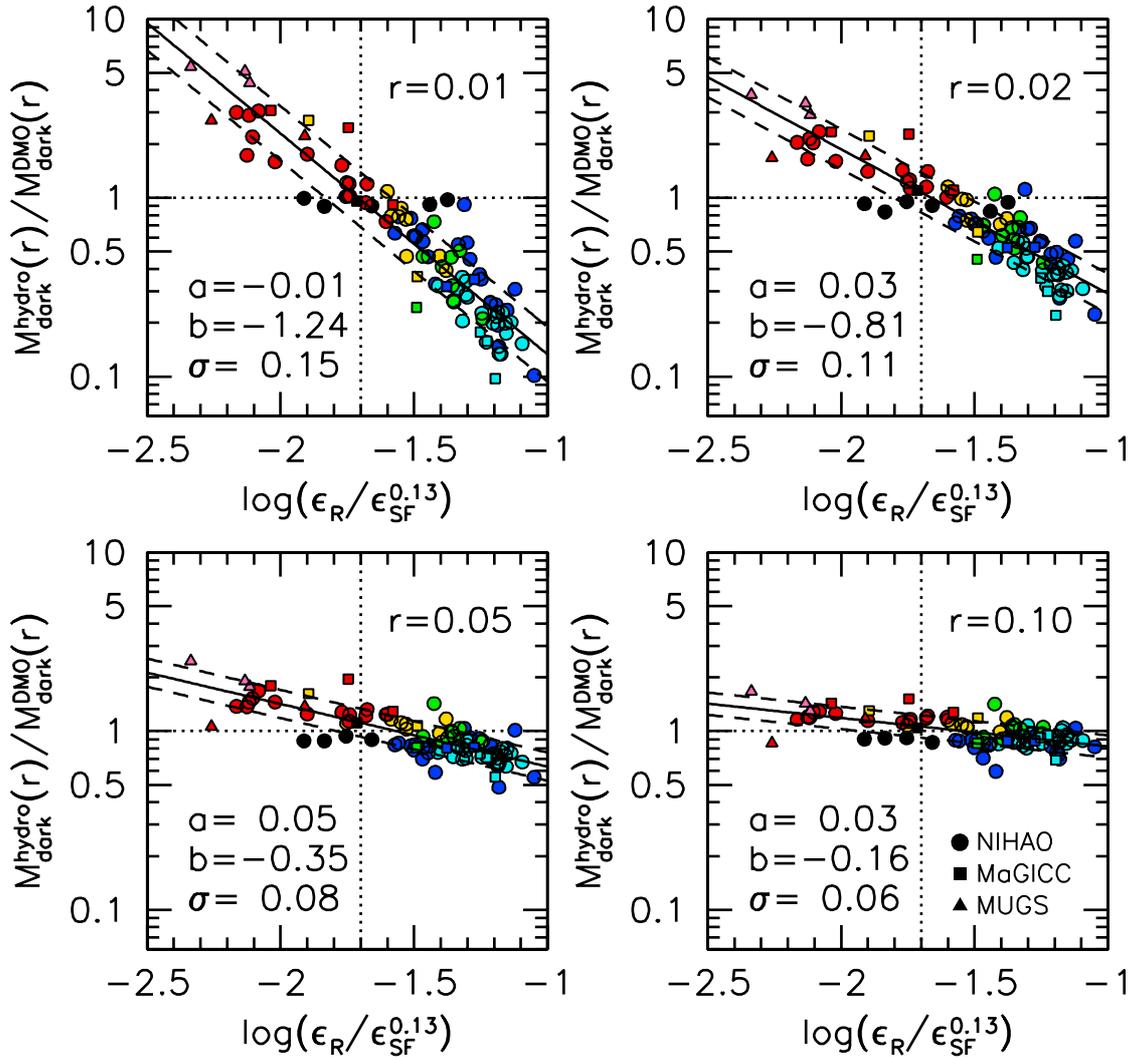,width=0.85\textwidth} }
  \caption{Halo response at 1, 2, 5, and 10 per cent of the virial
    radius versus $\eR/\eSF^{0.13}$. At all radii, the relation is
    well fitted with a power law. Point types are as in previous
    figures: NIHAO (circles); MaGICC (squares); MUGS (triangles).}
\label{fig:fp}
\end{figure*}

\subsection{Halo response as a function of star formation efficiency and galaxy compactness}

To determine the relative importance of star formation efficiency, and
galaxy compactness, Fig.~\ref{fig:rm} shows correlations between size,
stellar mass, and halo response. Here, the halo response is defined as
the dark matter mass ratio at 1 per cent of the virial radius.  The
left-hand panel shows the relation between $\eR$ and $\eSF$ with
points colour coded according to the halo response. The right-hand
panel shows the directly observable (2D half-light) size versus
stellar mass.  In both panels, we show the MW using a disk scalelength
of $R_{\rm d}=2.15$ ($\rhalf\sim 3.6$) kpc and $\Mstar=4.6\times
10^{10}\Msun$ from \citet{BovyRix13} and a halo mass of
$M_{200}=10^{12}\Msun \pm 0.2$ dex \citep[][and references
  therein]{Penarrubia16}.  In the left-hand panel, we show
$\rhalf=0.015R_{200} \pm 0.2$ dex derived by \citet{Kravtsov13} using
halo abundance matching. In the right-hand panel, we show
observations of 2D half-light radii for red, blue, and dSph galaxies
from \citet{Baldry12}.
    
Interestingly, galaxies with similar halo response fall in similar
parts of the parameter space.  As we saw in Fig.~\ref{fig:vv3} at
fixed $\eSF$, there is a clear trend for larger galaxies
(at fixed star formation efficiency or mass) to have more expanded
haloes, and smaller galaxies to have more contracted haloes.  However,
at a fixed $\eR$ (or size) there are two branches with
different $\eSF$ (or mass) for the same halo
response. Thus neither compactness nor star formation efficiency alone
determines the halo response.

If we fit lines to haloes of similar response where there is a
significant baseline, we find slopes of $\sim 0.13$ for \eR versus \eSF
and $\sim 0.2$ for $R_{V}^{\rm 2D}$ versus $\Mstar$.  Constant surface
density [$\Sigma=\Mstar/(2\pi R_V^2)$] has a slope of 0.5 in the $R_V$
versus \Mstar plane, and thus the density of stars is not a more
fundamental parameter.  This is clearly seen for dwarf galaxies, which
have roughly constant surface density of $\Sigma=2 \Msun\pc^{-2}$, while the
halo response varies from no change at low stellar masses ($\sim
10^6\Msun$) to expansion at high stellar masses $(\sim 10^8 \Msun)$.

\begin{figure}
  \centerline{
    \psfig{figure=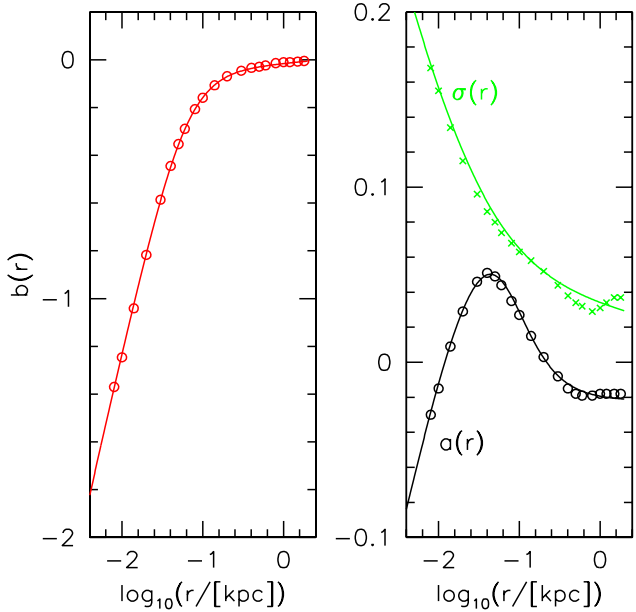,width=0.48\textwidth} }
  \caption{The halo response, $M_{\rm dark}^{\rm hydro}/M_{\rm
      dark}^{\rm DMO}(r)$, at each radius, $r$, is fitted with
    equation(\ref{eq:haloresponse}). The left-hand panel shows the
    slope, $b(r)$, while the right-hand panel shows the normalization,
    $a(r)$, and scatter, $\sigma(r)$. The solid lines show our fitting
    formula.}
\label{fig:fp2}
\end{figure}

Taking out these trends, Fig.~\ref{fig:fp} shows the halo response
measured at various radii versus $\eR/\eSF^{0.13}$.  At all radii,
$r$,  the relation is well fitted with a power law in $M_{\rm
  dark}^{\rm hydro}(r)/M_{\rm dark}^{\rm DMO}(r)$ versus
$\eR/\eSF^{0.13}$ space. The slope, $b$, zero-point, $a$,  and
scatter, $\sigma$, of this correlation depends on the radius where the
mass ratio is measured, $r$.  Thus, the following equation describes
the halo response at any radius as a function of $\eR$ and $\eSF$:
\begin{equation}
\label{eq:haloresponse}
  \log_{10}\left[\frac{M_{\rm dark}^{\rm hydro}(r)}{M_{\rm dark}^{\rm DMO}(r)}\right] = a(r) + b(r)[\log_{10}\left(\eR/\eSF^{0.13}\right) -1.7 ].
\end{equation}
The dependence of $a$, $b$, and $\sigma$, with radius are shown in Fig.~\ref{fig:fp2}. We fit the relations with the following functions.
A double power-law with slopes $s_1$ and $s_2$:
\begin{eqnarray}
  D(x, x_0, y_0, s_1, s_2, t) = y_0  + s_1(x-x_0) \nonumber \\
  +\frac{(s_2-s_1)}{t} \log_{10}\left\{\frac{1}{2} +\frac{1}{2}\left[10^{(x-x_0)}\right]^{t}\right\},
\end{eqnarray}
and a sigmoid function which smoothly varies between a $y$-value of $y_1$ and $y_2$:
\begin{equation}
 S(x, x_0, y_1, y_2, t) = y_2 +\frac{y_1-y_2}{1 + (10^{(x-x_0)})^t}.
\end{equation}
The slope as a function of radius, $b(r)$, is given by
\begin{equation}
b(r) = D\left[\log_{10}(r), -1.1, -0.2, 1.35, 0.0, 2.0\right],
\end{equation}
the zero-point as a function of radius, $a(r)$, is given by
\begin{eqnarray}
  a(r) = D\left[\log_{10}(r), -0.15, 0.085, 1.5, 0.0, 2.0\right] \nonumber \\ 
  +S\left[\log_{10}(r). -1.0, 0.0,-0.13, 2.0\right]
\end{eqnarray}
and the 1$\sigma$ scatter is given by
\begin{equation}
\label{eq:hrscatter}  
\sigma(r) = D(\log_{10}(r),-1.1, 0.65,-1.4,0.0,1.0).
\end{equation}

These predictions break down at low star formation efficiencies
($\eSF\lta 0.001$), so we apply an S-function to the halo response
(equation~\ref{eq:haloresponse}) and the scatter (equation~\ref{eq:hrscatter})
with $S[\log_{10}(\eSF),-3,0,1,2]$. This function asymptotes to zero as $\eSF
\rightarrow 0$, so we add in quadrature a scatter of 0.03 dex.

\begin{figure*}
    \centerline{
    \psfig{figure=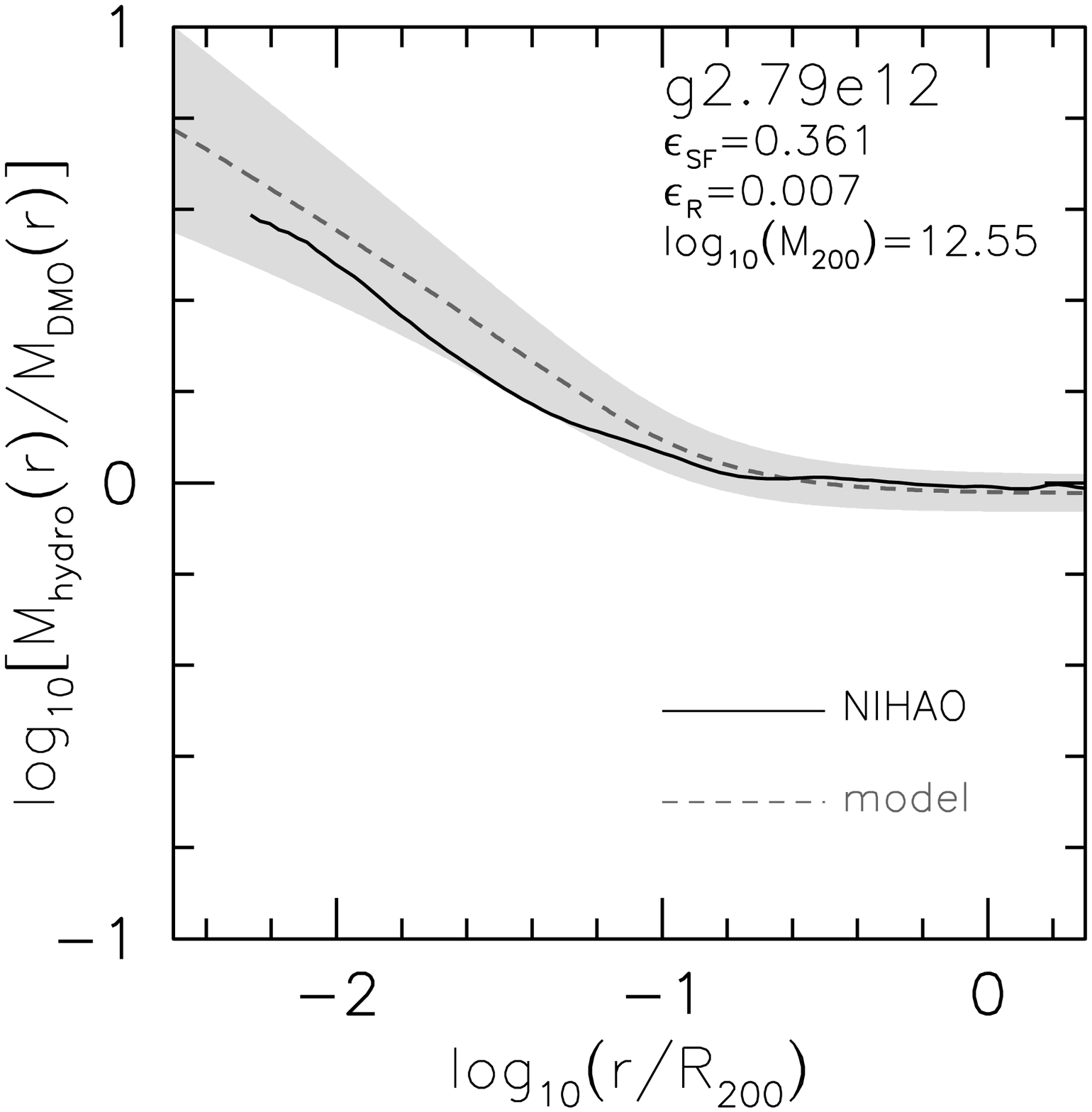,width=0.24\textwidth}
    \psfig{figure=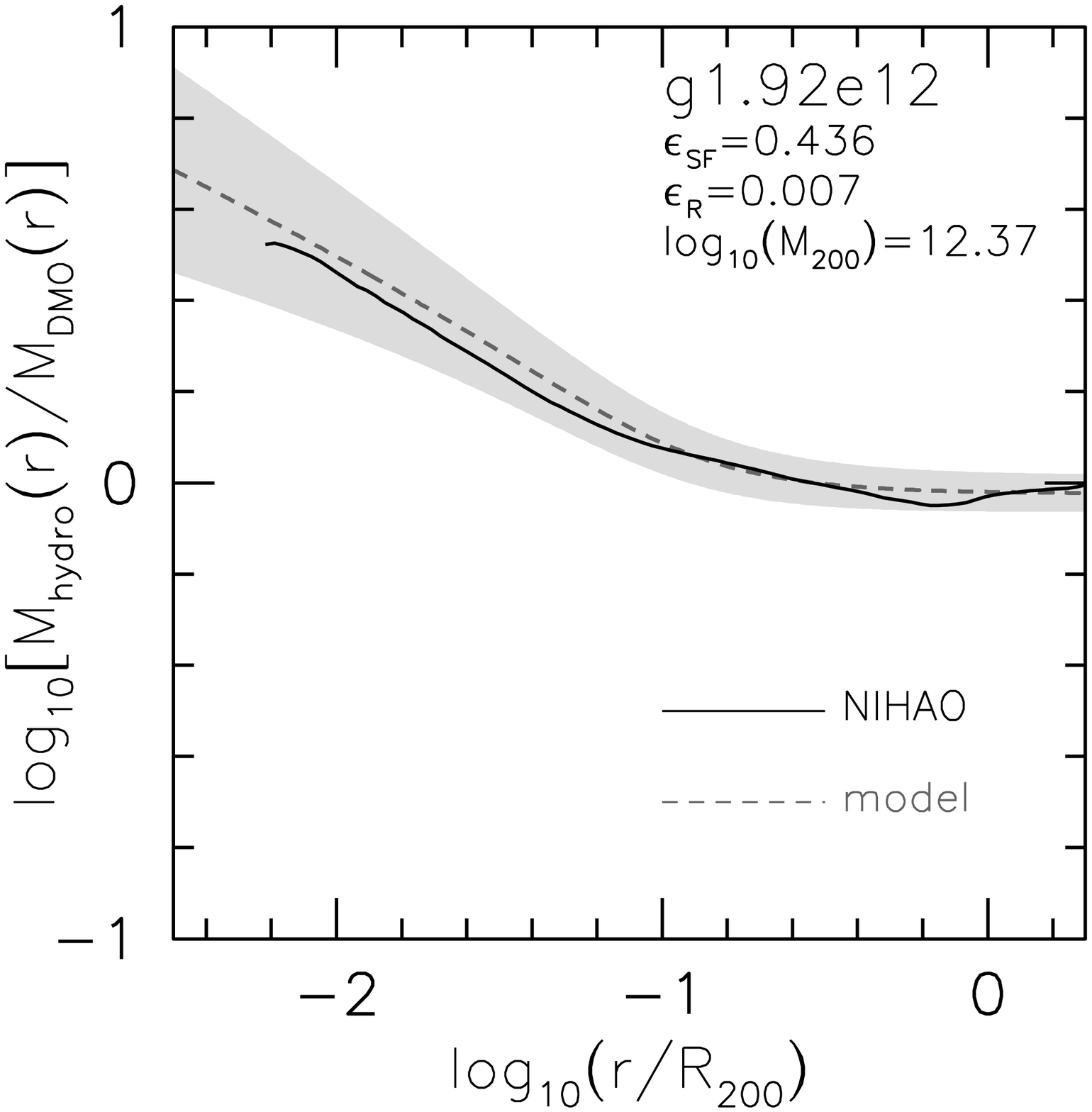,width=0.24\textwidth}
    \psfig{figure=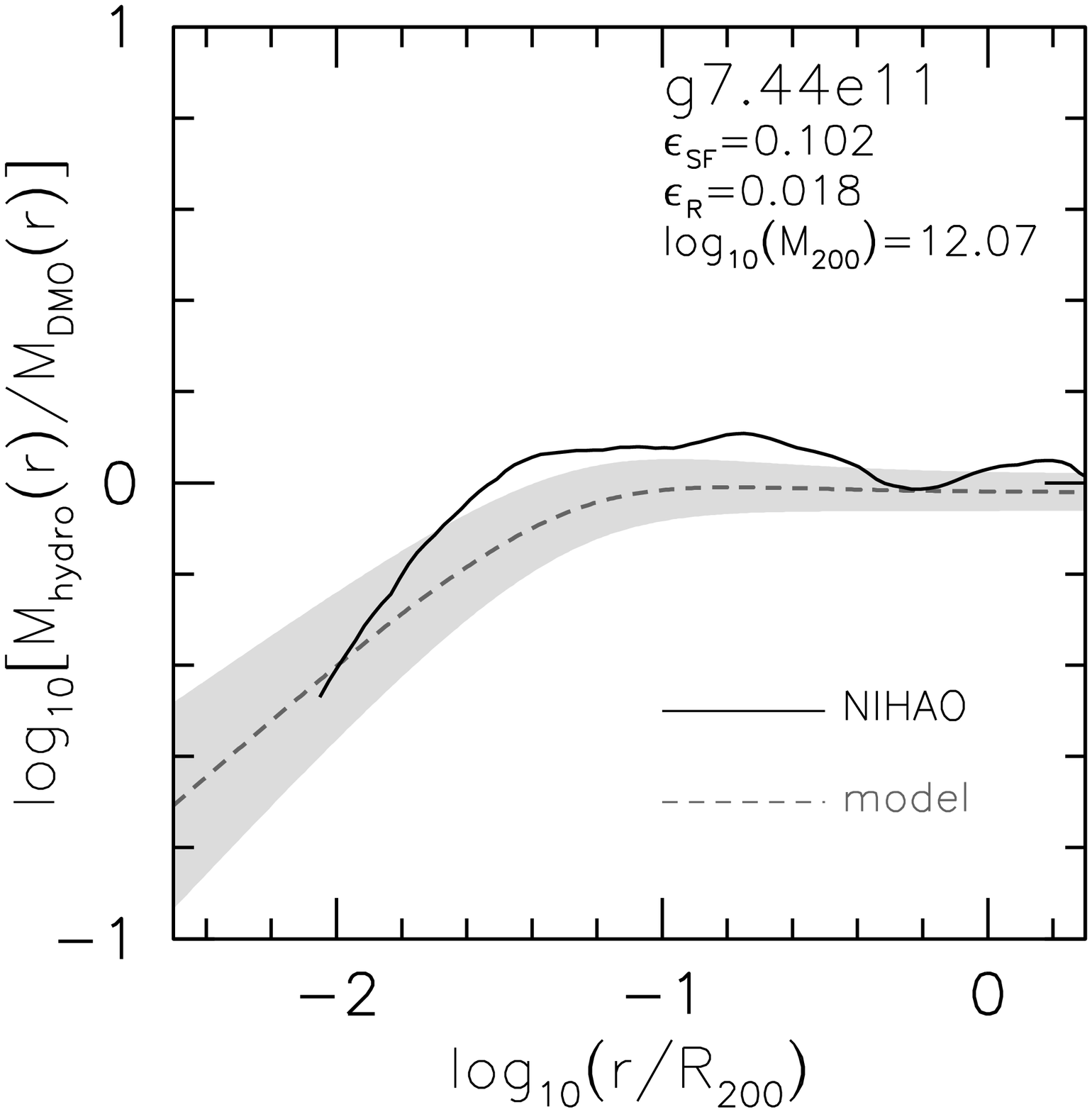,width=0.24\textwidth}
    \psfig{figure=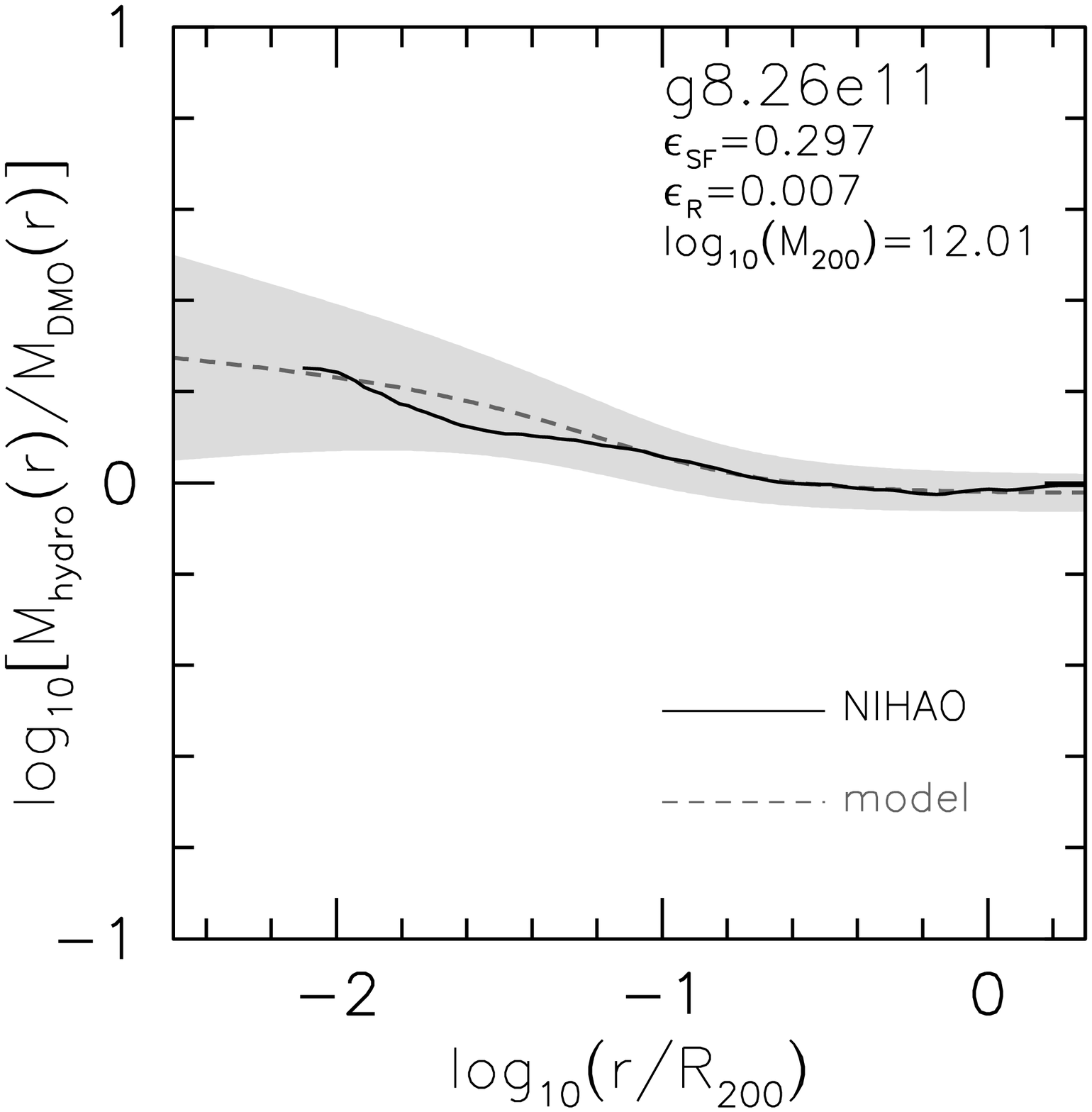,width=0.24\textwidth}
    }
  \centerline{
    \psfig{figure=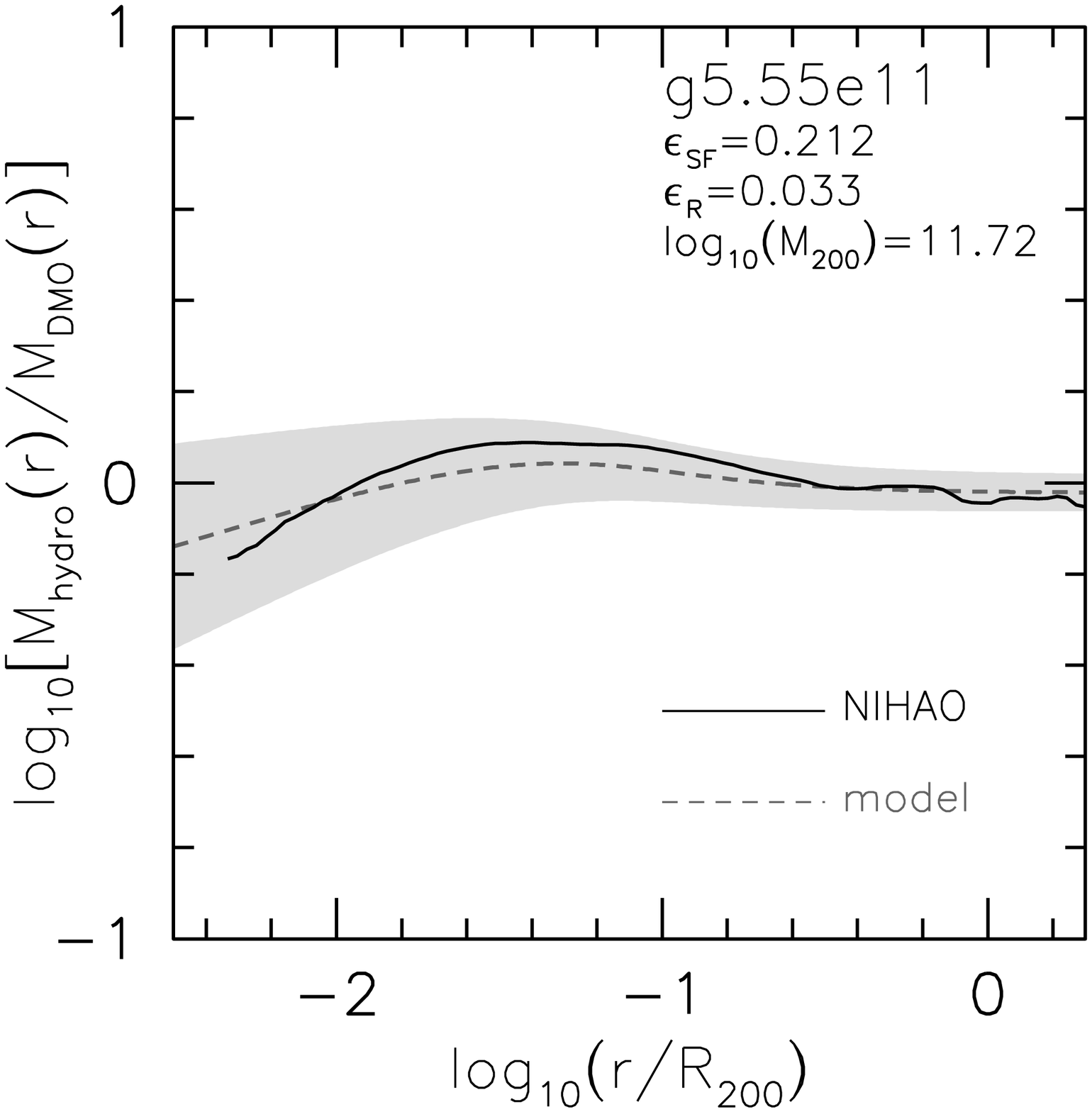,width=0.24\textwidth}
    \psfig{figure=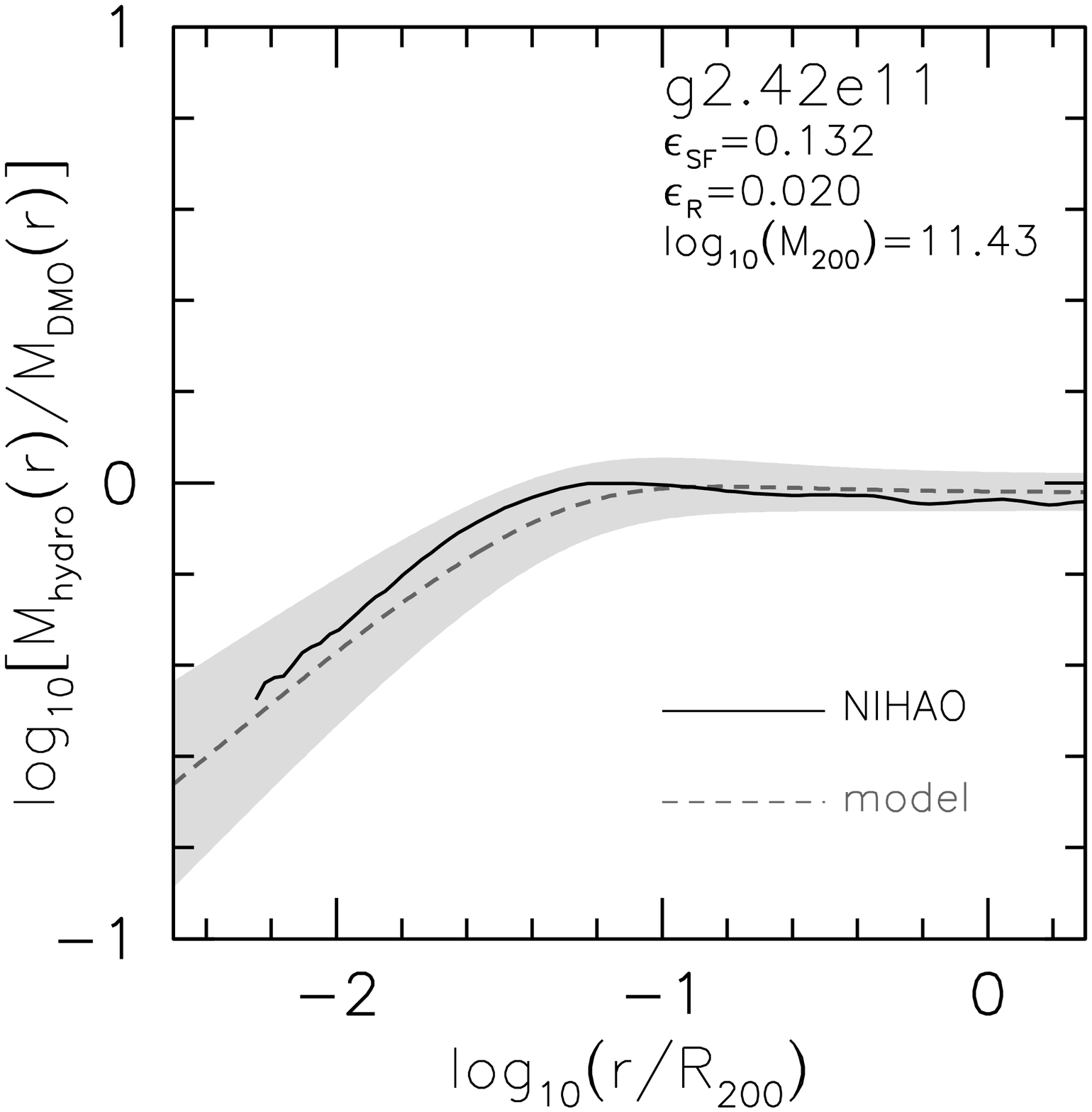,width=0.24\textwidth}
    \psfig{figure=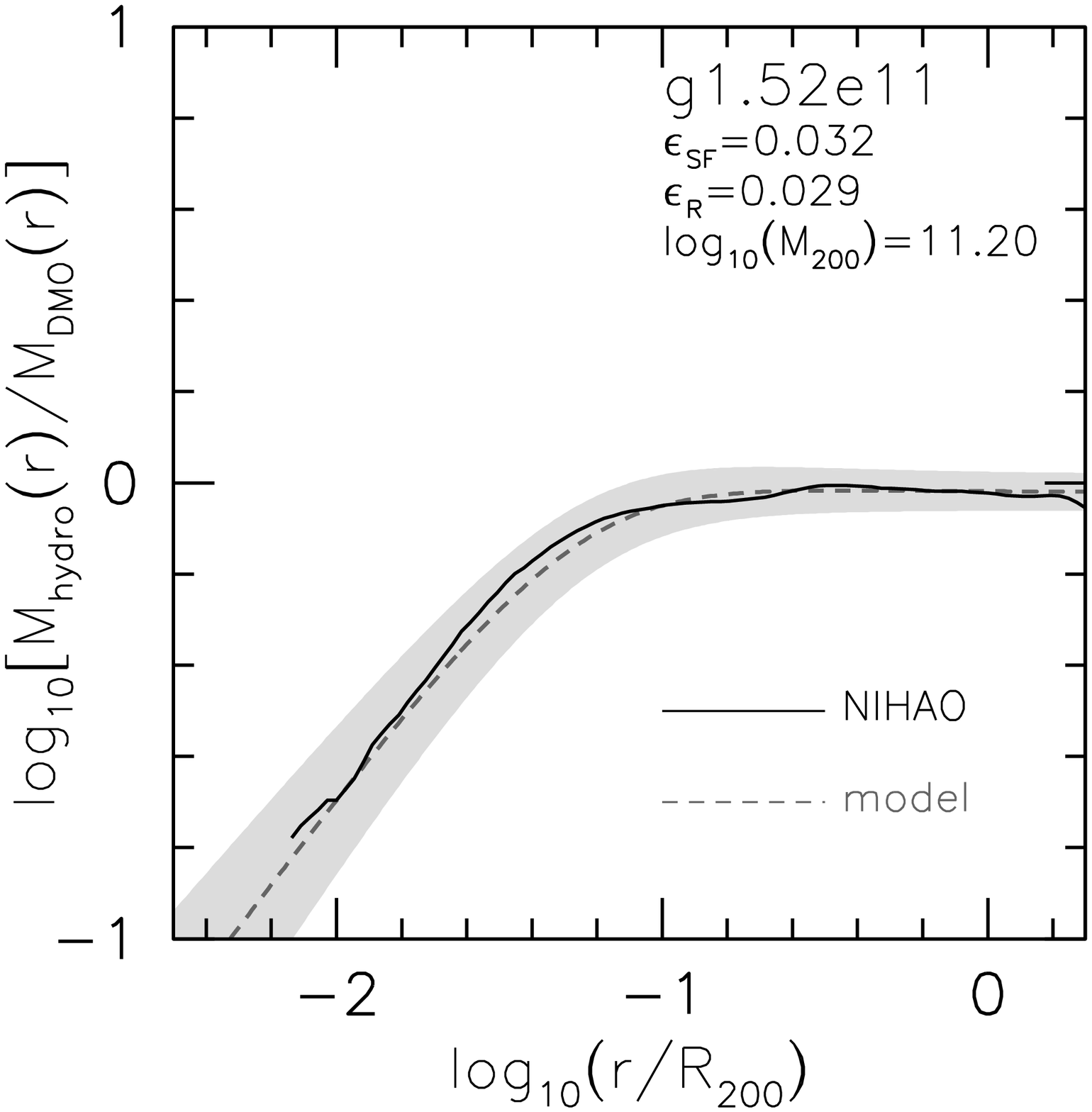,width=0.24\textwidth}
    \psfig{figure=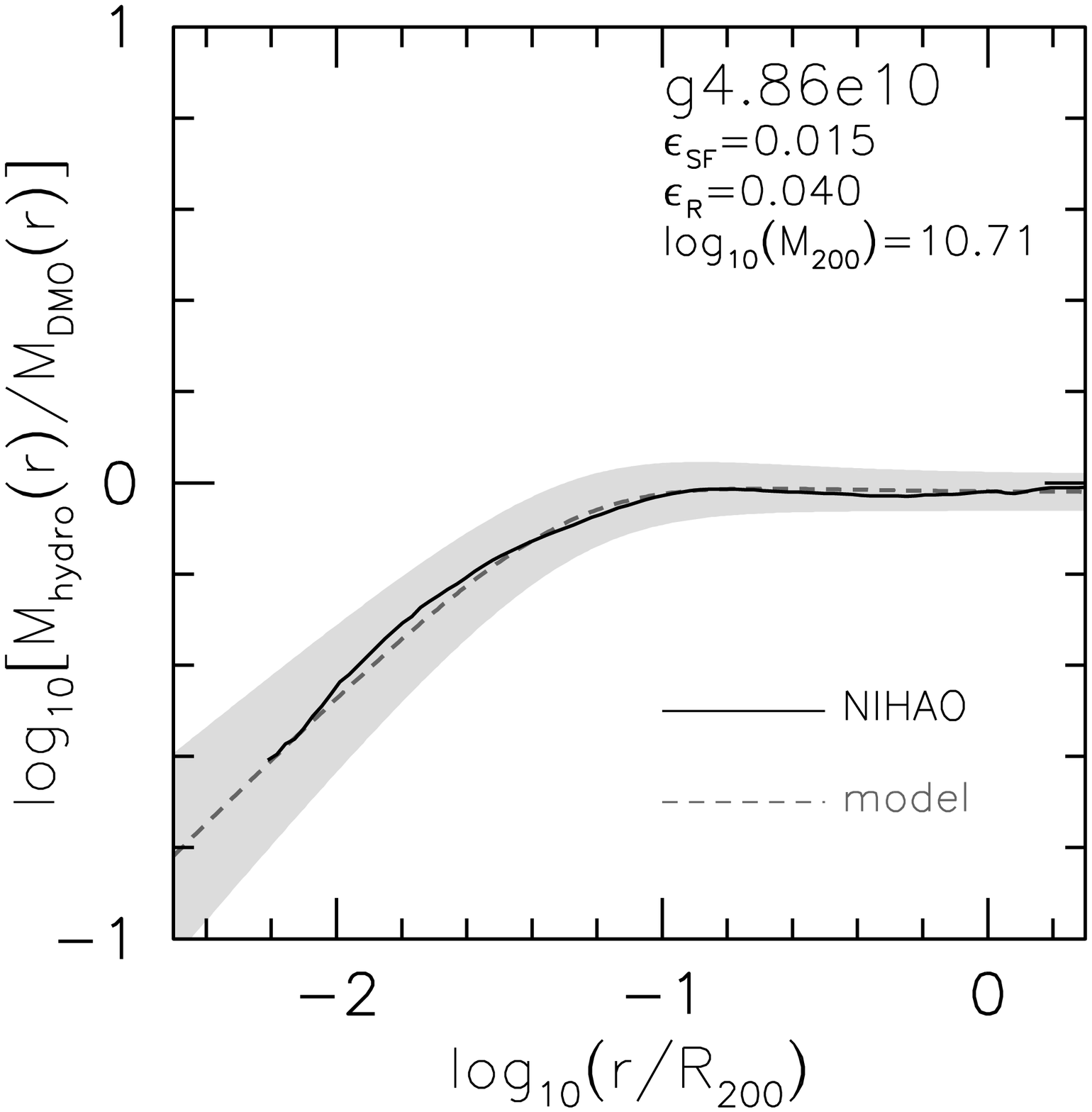,width=0.24\textwidth}
  }
  \centerline{
    \psfig{figure=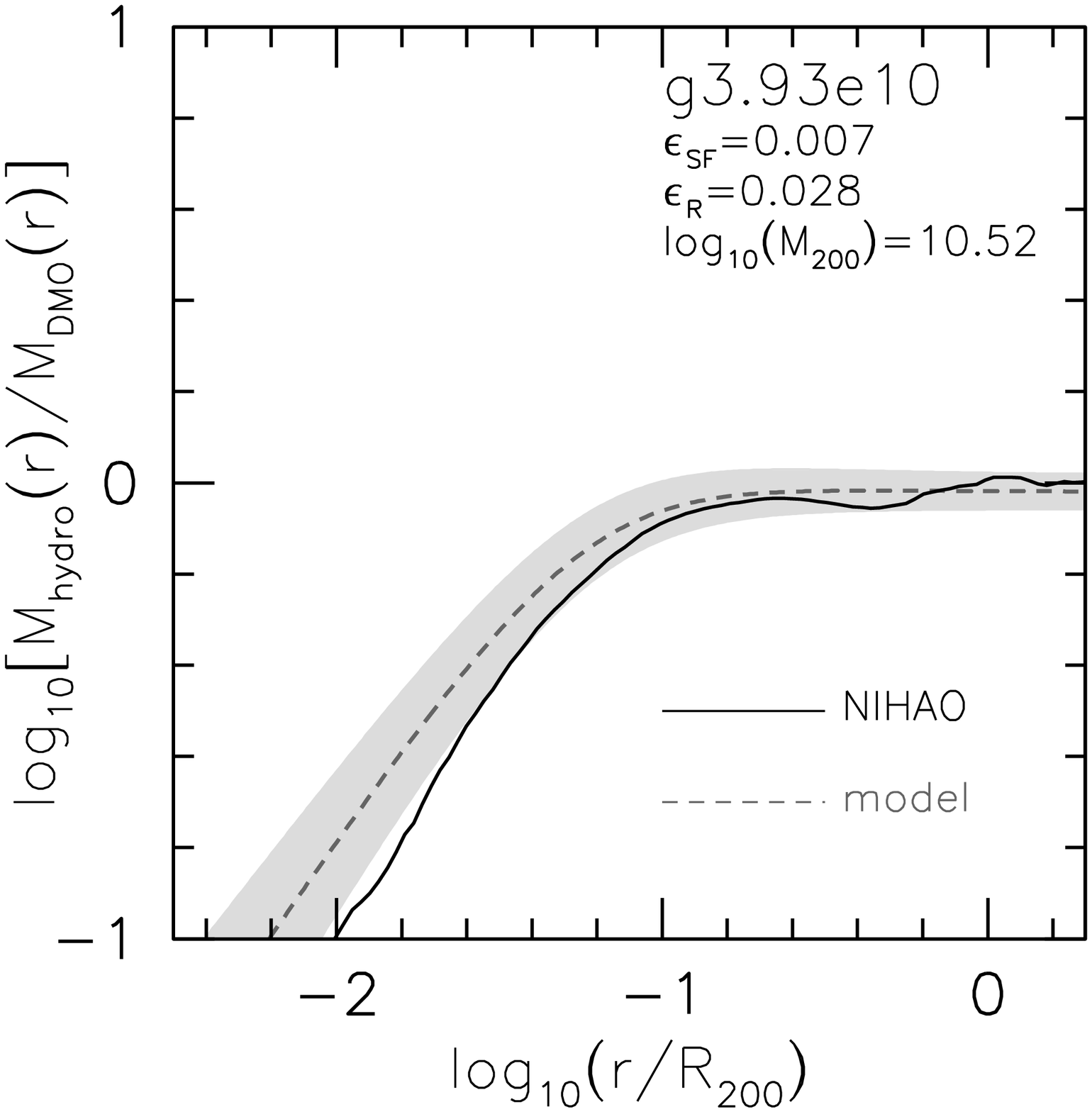,width=0.24\textwidth}
    \psfig{figure=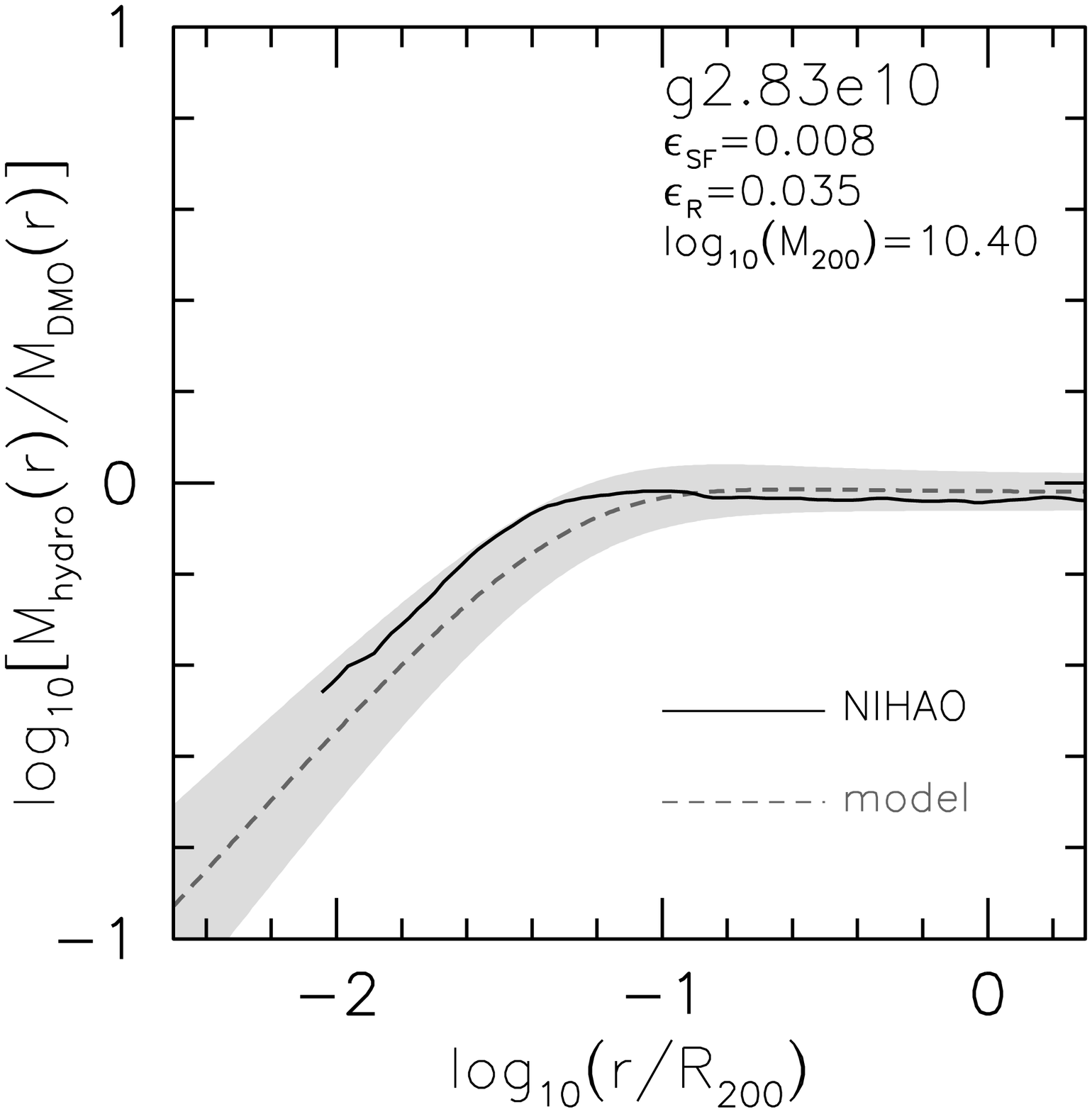,width=0.24\textwidth}
    \psfig{figure=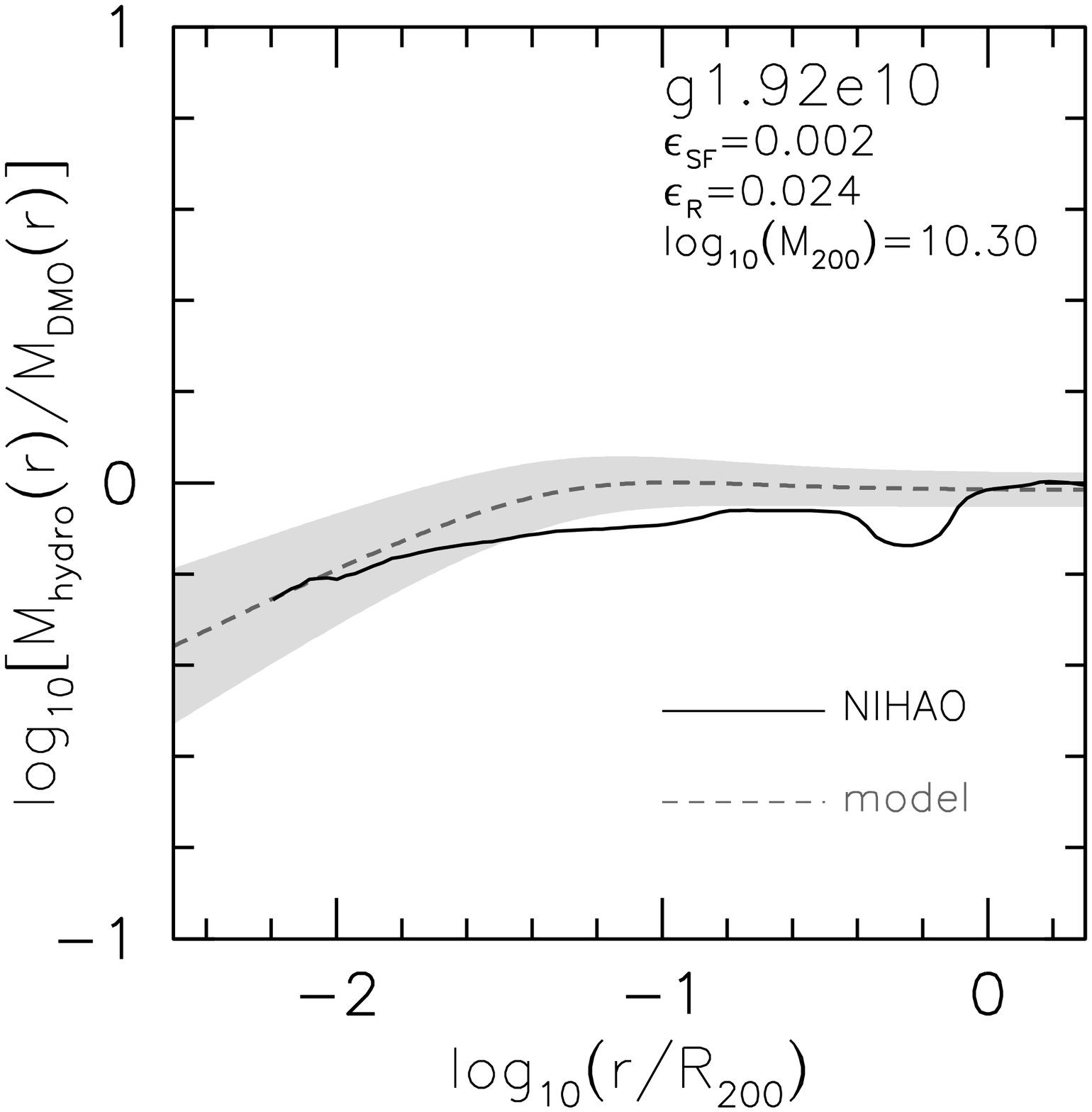,width=0.24\textwidth}
    \psfig{figure=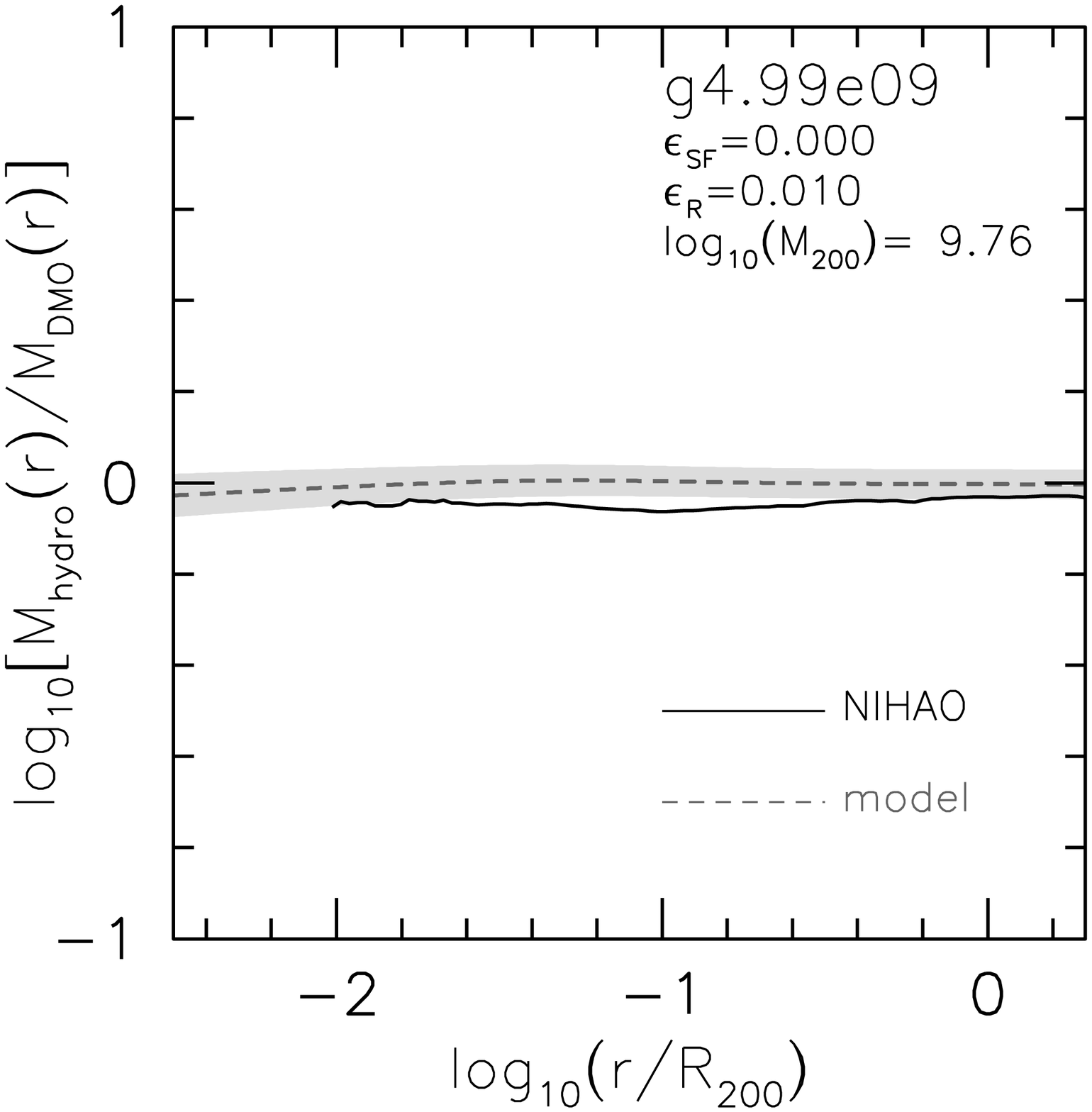,width=0.24\textwidth}
  }
  \caption{Predicted change (dashed lines) with $1\sigma$ uncertainty
    (shaded region) versus actual change in dark halo mass profiles (solid
    black lines) for a representative subset of NIHAO galaxies
    ordered by halo mass from upper left to lower right. In each panel,
    the galaxy ID, star formation efficiency, compactness parameter,
    and halo masses are given.}
\label{fig:xmm2}
\end{figure*}

Fig.~\ref{fig:xmm2} shows a comparison between the halo response in
the NIHAO simulations (solid black lines) and that predicted by our
analytical fit (dashed line and shaded region) for a representative
sub-sample of haloes. Reassuringly the model, which was calibrated
against these simulations, does indeed recover the diversity of halo
responses.

\subsection{Implications}
\label{sec:implications}
If this halo response model holds outside the range of \eR and \eSF
where we have simulations, then there are some interesting
implications.  Fig.~\ref{fig:rm} shows that our simulations do not
cover the full range of observed galaxies in the size versus mass plane.
At stellar masses between $\sim 10^7$ and $10^9\Msun$ our simulations
lack small galaxies, while at high masses $\sim 10^{11}\Msun$ our
simulations lack large galaxies.  Part of this discrepancy is due to
low mass red galaxies preferentially being satellites, whereas all the
NIHAO galaxies we show are centrals.  Comparing to the just observed
blue galaxies, the NIHAO simulations are offset high by $\sim 1\sigma$
at a stellar mass of $\sim 10^8\Msun$, and offset low by $\sim
1\sigma$ at a stellar mass of $\sim 10^{10}\Msun$. The small half-light
sizes can be traced to overly compact bulges, rather than the discs being too
small.  We note that a similar feature is seen in the EAGLE
simulations \citep{Crain15} with constant stellar feedback efficiency
(as we adopt here). Their solution is to increase the efficiency of
the stellar feedback in dense gas, and to decrease the efficiency in
low-density gas, to correct for numerical radiative energy loses.
Even though our massive galaxies are smaller than typical observed
galaxies, they actually overlap with the MW, which is known to
be atypically small \citep{Hammer07}.

The slope of the size-mass relation is roughly the same as the slope
of constant halo response, i.e., $\sim 0.2$. Thus, we would predict
that a typical spiral galaxy (with $\Mstar \gta 10^8 \Msun$) would
have mild halo expansion, a galaxy offset by $1\sigma$ to large sizes
would have strong expansion, while a $1\sigma$ small galaxy (e.g.,
the MW) would have no change.  A typical bulge-dominated galaxy
would also have no change in the dark matter halo.  This is in
contrast to the strong mass dependence found in the NIHAO and MaGICC
simulations here and in previous works \citep{DiCintio14a,
  DiCintio14b, Tollet16}. The strong mass dependence can be traced to
the strong dependence of compactness with star formation efficiency. 

\setcounter{figure}{8}


\section{Towards a physical model for halo response}
\label{sec:physicalmodel}
We now turn our attention to a physical model for the response of CDM
haloes to baryonic galaxy formation processes.  A more elaborate
analysis of this toy model is given in Dekel \etal (in preparation).

\subsection{Adiabatic contraction formalism}

Given a spherically enclosed mass profile from a DMO simulation,
$\Mi(r)$, (where the $\rm i$ refers to initial), we wish to derive the
final dark matter profile, $M_{\rm dm,f}(r)$, once the final baryonic
mass profile, $M_{\rm b,f}(r)$, is specified. It is assumed the
initial mass profile is split into baryons and dark matter according
to the cosmic baryon fraction: $M_{\rm b,i}(r)=\fbar \Mi(r)$, and
$M_{\rm dm,i}(r)=(1-\fbar) \Mi(r)$.

The standard model of halo response is adiabatic contraction
introduced by \citet{Blumenthal86}.  As well as adiabatic invariance,
this model assumes spherical symmetry, homologous contraction, and
circular particle orbits, resulting in conservation of the angular
momentum and thus $r M(r) =const$, where $M(r)$ is the total mass
enclosed within radius $r$.  Additionally assuming the dark matter
shells do not cross: $M_{\rm dm,f}(\rf)=M_{\rm dm,i}(\ri)$, where \ri
is the `initial' radius of a shell of dark matter, and \rf is the
`final' radius of this shell after the effects of baryonic processes
are included, yields
\begin{equation}
\label{eq:ac}
\rf/\ri = \Mi(\ri)/[M_{\rm b,f}(\rf)+\Mi(\ri)(1-\fbar)].
\end{equation}
Thus given $\Mi$ and $M_{\rm b,f}$, one can solve equation (\ref{eq:ac}) for
the mapping between $\rf$ and $\ri$, and hence derive the final dark
matter profile.

\begin{figure}
\centerline{
\psfig{figure=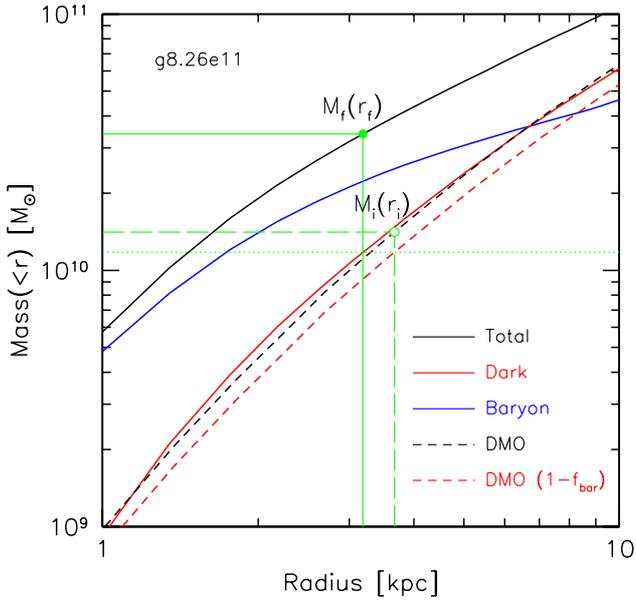,width=0.48\textwidth}
}
\caption{Example showing how $\rf, \ri, \Mf, \Mi$ are calculated in (a
  portion of) the cumulative mass vs radius plot for simulation
  g8.26e11. The horizontal green dotted lines shows an arbitrary mass,
  here $10^{10.07}\Msun$. The radius this intersects the dark matter
  profile from the hydro simulation is termed $\rf$ (vertical green
  solid), while the radius this intersects the dark matter profile
  from the DMO simulation is termed $\ri$ (vertical green
  dashed). The total masses contained within $\rf$ and $\ri$ are then
  given by $\Mf$ (horizontal green solid) and $\Mi$ (horizontal green
  dashed) respectively.}
\label{fig:rfi}
\end{figure}

An example of how the radii and masses are calculated in one of our
simulations is shown in Fig.~\ref{fig:rfi}. The solid lines show the
total (black), dark matter (red), and baryonic (blue) mass profiles
for the hydro simulation for halo g8.26e11. The dashed lines show the
total (black) and implied dark matter (red) mass profiles for the
corresponding dark matter only (DMO) simulation. The horizontal dotted
green line shows an arbitrary mass, which intersects the hydro
(`final') dark matter mass profile at $\rf$ and the DMO
(`initial') dark matter mass profile at $\ri$. \Mf and \Mi are then
the total enclosed mass at \rf and \ri, respectively. The arbitrary
mass is varied from the smallest resolved to the virial mass to obtain
a profile of $\rf/\ri$ and $\Mi/\Mf$.  We can then compare the
relation between $\rf/\ri$ and $\Mi/\Mf$ from the simulations to test
the adiabatic contraction model.

\subsection{Halo response in NIHAO simulations}
\begin{figure}
\centerline{
\psfig{figure=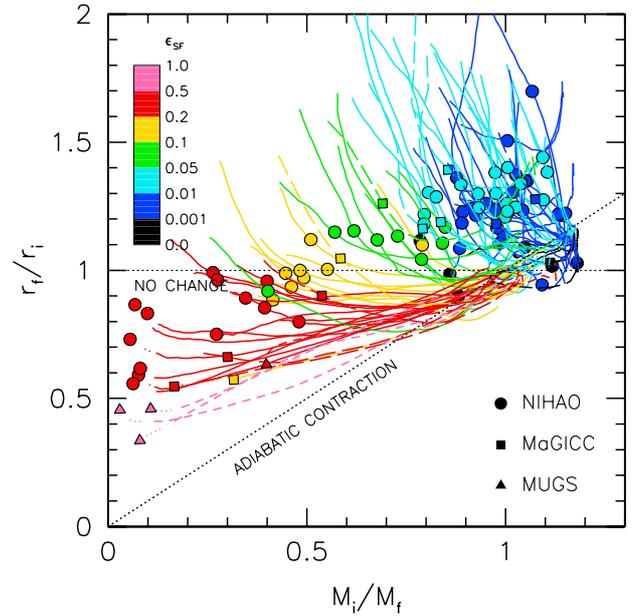,width=0.48\textwidth}}
\caption{Halo response is not adiabatic, nor does it follow a
  universal path in the radius versus mass plane. Rather, the path is
  correlated with the star formation efficiency, $\eSF$.
  Each line shows a different dark matter halo. Lines are colour coded
  by star formation efficiency (as indicated).}
  \label{fig:rrmm}
\end{figure}

Fig.~\ref{fig:rrmm} shows the tracks of our simulations in the
$\rf/\ri$ versus $\Mi/\Mf$ plane. The results from each simulation pair
are colour coded by the star formation efficiency, and are plotted down
to the convergence radius of the DMO simulation. The
$y$-axis shows the `contraction' factor, while the $x$-axis can
(typically) be mapped monotonically to radius. The adiabatic
contraction formula (Eq.~\ref{eq:ac}) predicts that $\rf/\ri=\Mi/\Mf$,
which is indicated by the diagonal dotted line in the figure. No
change in the dark matter profile corresponds to the horizontal dotted
line at $\rf/\ri=1$.

\begin{figure*}
\centerline{
\psfig{figure=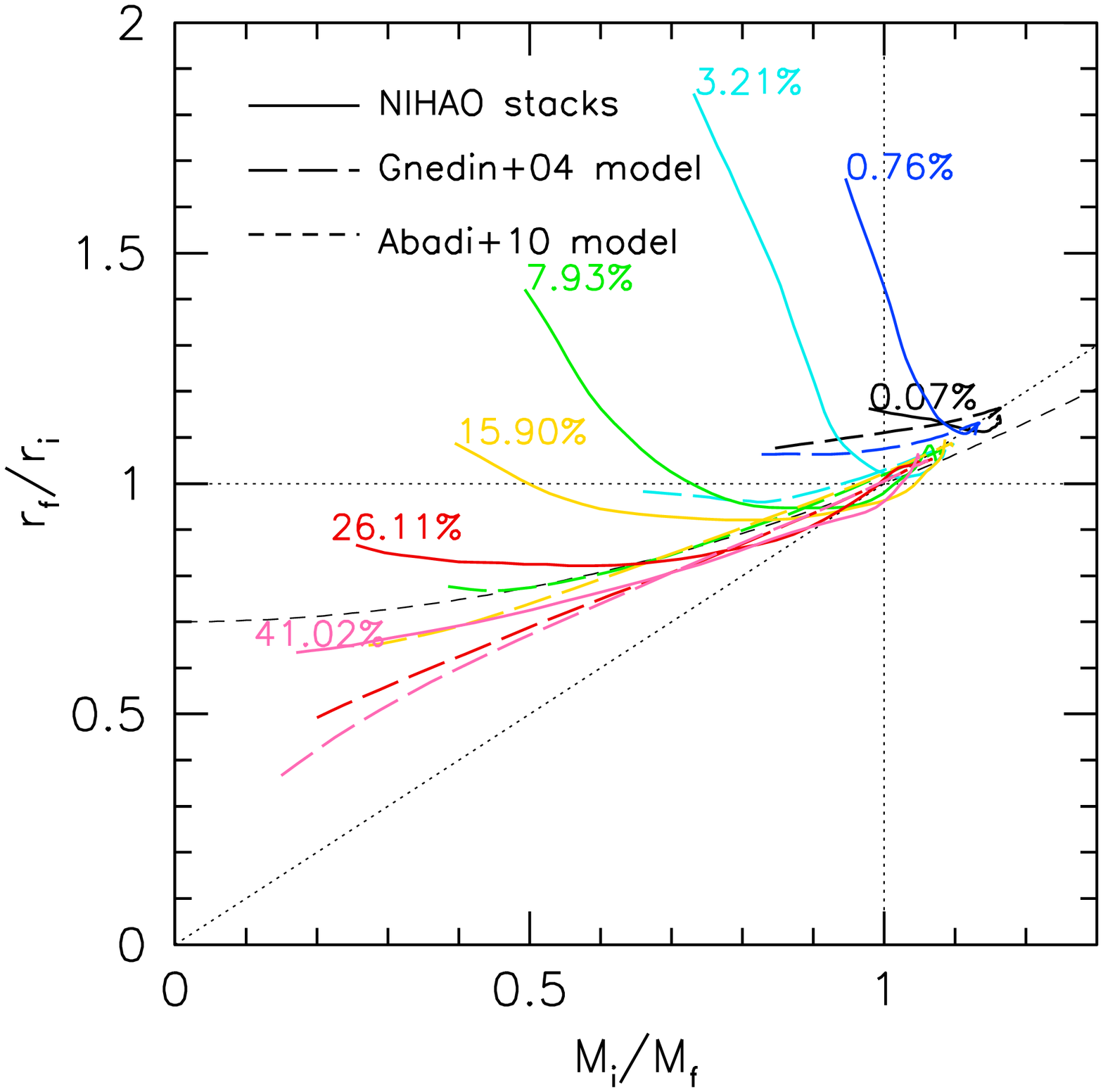,width=0.48\textwidth}
\psfig{figure=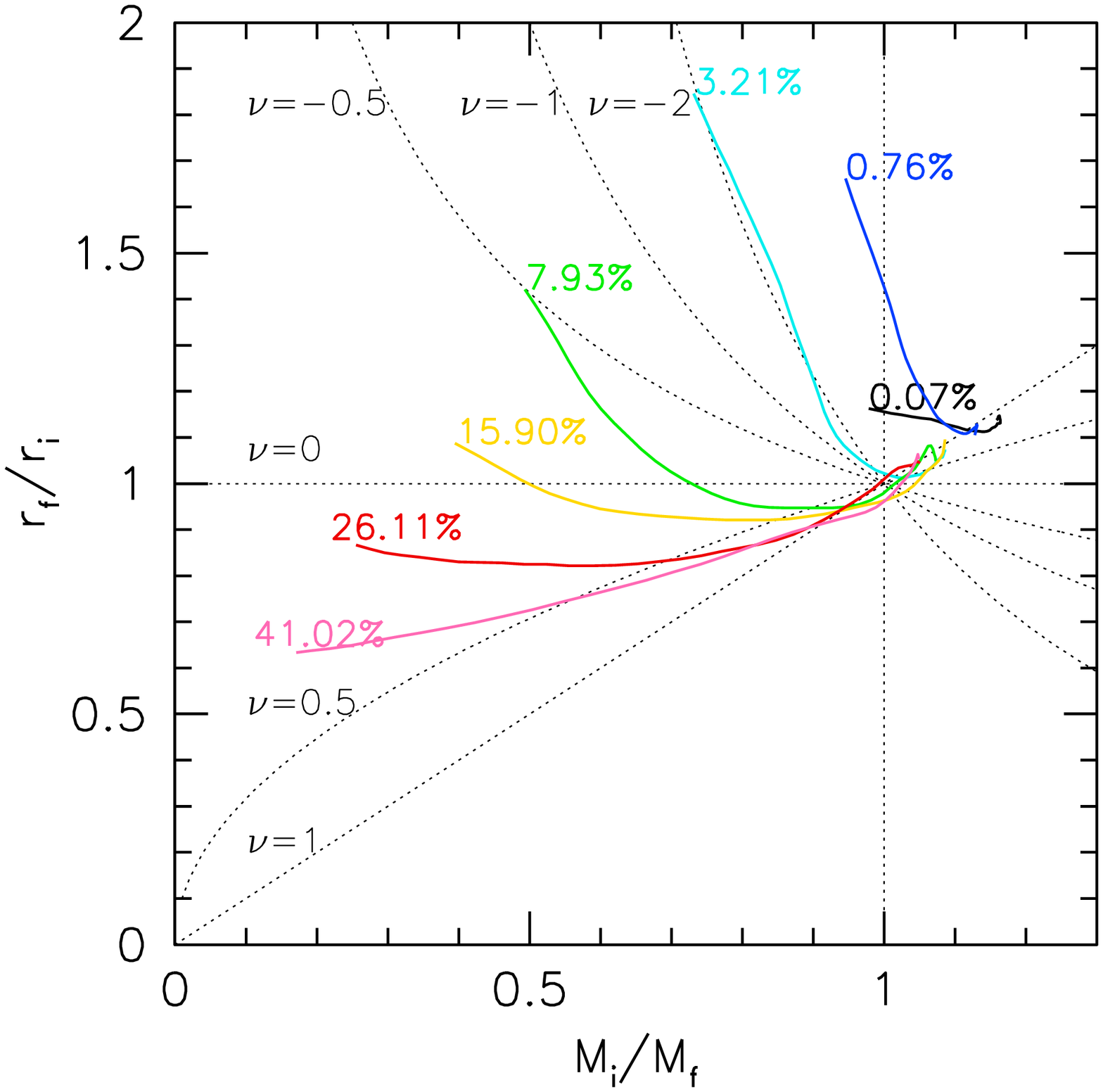,width=0.48\textwidth}}
\caption{Comparison between halo response in NIHAO simulations (binned
  by star formation efficiency, solid lines), and various models in
  the literature. Left: long-dashed lines show the halo response
  predicted by the model of \citet{Gnedin04}, which is different for
  each NIHAO stack, while the short dashed line shows the relation
  from \citet{Abadi10}. Right: dotted lines show
  $\rf/\ri=(\Mi/\Mf)^{\nu}$ which is sometimes used to parametrize the
  halo response in galaxy mass models. None of these models is able to
  capture the full variation in halo response.}
  \label{fig:rrmm2}
\end{figure*}

The first point that is apparent from Fig.~\ref{fig:rrmm} is that the
halo response is not adiabatic in any of our simulations, confirming
previous studies \citep{Gnedin04, Abadi10, Pedrosa10, Dutton15}.  The
second point is that the halo response does not follow a single track.
However, in contrast to these studies, which found only contraction
with a mild variation in the halo response, our simulations show a
much wider diversity of halo responses, including expansion.  As
expected from Fig.~\ref{fig:vv2} the halo response is strongly
correlated with the star formation efficiency.  This is more clearly
shown in Fig.~\ref{fig:rrmm2}, which shows the halo response for NIHAO
haloes (solid lines) binned by star formation efficiency. Here, the
lines are plotted from 0.01 $R_{200}$ to $R_{200}$.

The deviation from adiabatic contraction is increasing with decreasing
$\eSF$: haloes with lower star formation efficiency have higher
$\rf/\ri$ (i.e., less contraction / more expansion) at fixed
$\Mi/\Mf$.  The long-dashed lines in the left panel of
Fig.~\ref{fig:rrmm2} show the halo response predicted by the
\citet{Gnedin04} model. This model results in weaker contraction for
lower $\eSF$ galaxies, and even mild expansion for very low $\eSF$,
however, for all star formation efficiencies this model over predicts
the contraction. The discrepancy is largest for low $\Mi/\Mf$ (i.e.,
small radii). The short-dashed lines show the model of
\citet{Abadi10}, which has a fixed path in the halo response plane.
This model is close to our simulation results at $\eSF\sim 0.4$, but
fails to match lower $\eSF$ haloes.

The dotted lines in the right-hand panel show tracks of
$\rf/\ri=(\Mi/\Mf)^{\nu}$ for various values of $\nu$. This formula is
useful in semi-empirical galaxy formation models \citep{Dutton07} and
rotation curve mass models \citep{Dutton13a}, as with a single
parameter, $\nu$, one can smoothly vary between adiabatic contraction
($\nu=1$), no change ($\nu=0$), and expansion ($\nu < 0$). However,
despite its flexibility, this function does not capture the diversity
of halo response seen in our simulations. It only reliably recovers
the halo response for $\eSF\sim 0.03$, furthermore, it  lacks a direct
physical interpretation.  This failure of previous models and fitting
formulae motivates us to derive a new toy model for the halo response.

\subsection{Toy model for halo contraction and expansion}
\label{sec:toymodel}
We now discuss a toy model for halo response that will be useful for
guiding the discussion of the physical mechanisms at play.  We
emphasize here that we employ an isolated shell toy model, which is a
crude simplification. An analysis for a full halo density profile and the
error made by the shell approximation is given in Dekel \etal (in preparation).

\begin{figure*}
  \centerline{
    \psfig{figure=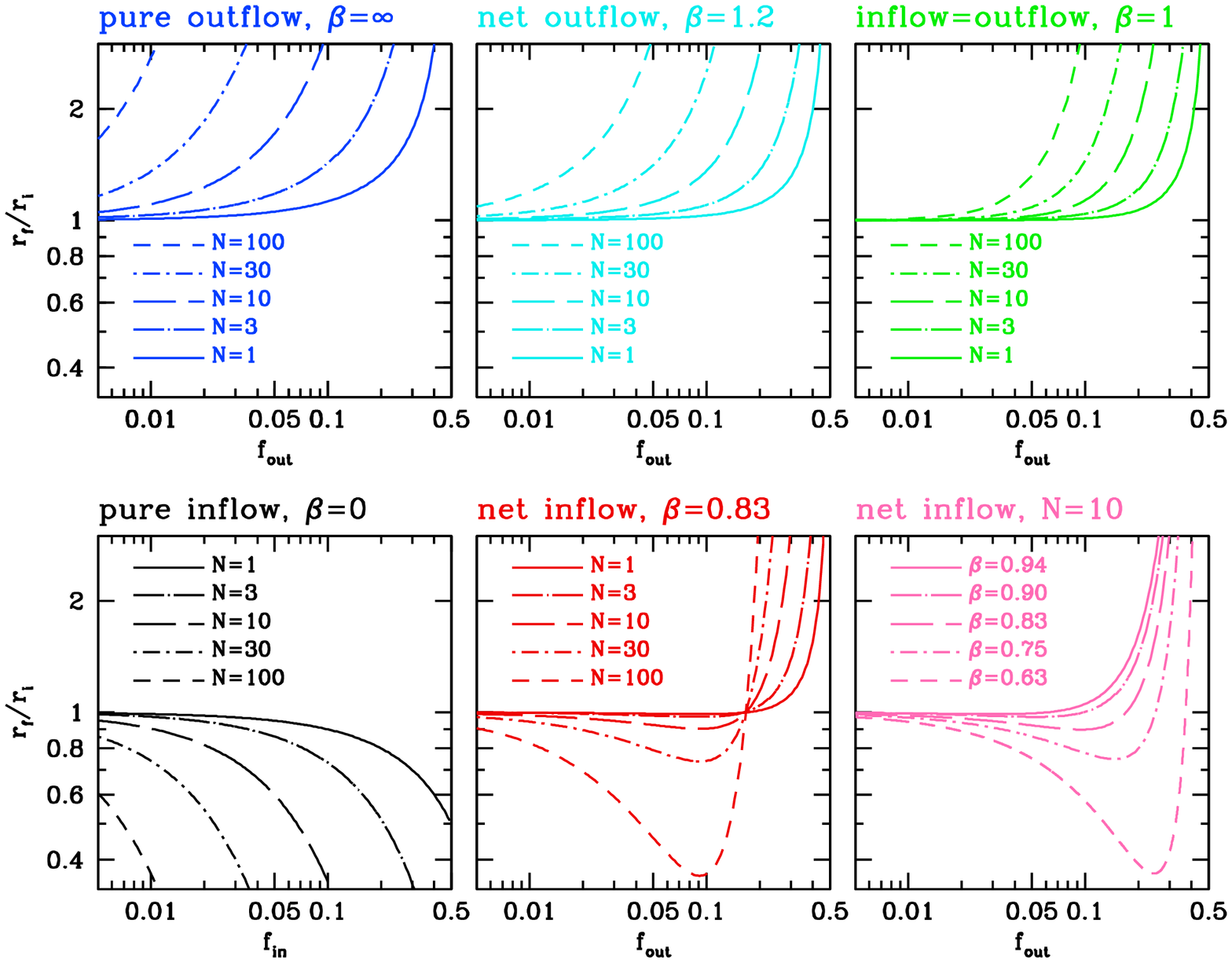,width=0.95\textwidth} }
  \caption{Toy model for adiabatic contraction and impulsive
    expansion. These plots show the change in radii of a shell,
    $\rf/\ri$, versus the fraction of mass in the adiabatic inflow
    ($\fin$) or impulsive outflow ($\fout$), with a ratio
    $\beta=\fout/fin$ and $N$ cycles.  Lower left: pure inflow
    ($\beta=0$) leads to contraction $(\rf/\ri<1)$. Upper left:
    pure outflow leads to expansion $(\rf/\ri > 1)$. For a single
    event, $N=1$, the halo response is only significant for $\fout\gta
    0.3$, while for multiple events even small values of $\fout$ can have
    large cumulative effects. Upper middle: net outflow with
    $\beta=1.2$ results weaker expansion than the pure inflow case.
    Upper right: equal inflow and outflow fractions ($\beta=1$),
    repeated $N$ times results in expansion for all values of
    $\fout$. Lower middle: when there is net inflow (e.g.,
    $\beta=0.83$) contraction occurs for $\fout < (1-\beta)$ and
    expansion for $\fout > (1-\beta)$. More cycles results in larger
    departures of $\rf\ri$ from unity. Lower right: net inflow
    with $N=10$ cycles and different $\beta$ can result in the same
    maximum contraction factors as variable $N$ at fixed $\beta$. }
  \label{fig:toymodel}
\end{figure*}

We consider two physical processes that change the dark halo
structure: {\it adiabatic inflow} of gas and {\it impulsive outflow}
of gas.  Fig.~\ref{fig:toymodel} shows a variety of cases:  pure
outflow (upper left); net outflow (upper middle); equal inflow and
outflow (upper right); pure inflow (lower left); and net inflow (lower
middle and right).  Fig.~\ref{fig:toymodel2} shows the paths of these
models, using the same colour scheme, in the classical halo response
plane: $\rf/\ri$ versus $\Mi/\Mf$.

\subsubsection{Adiabatic inflow}
Consider a spherical shell of mass $(1-\fin)M$ and radius $\ri$, with
circular orbits. Assume that gas of mass $\fin M$ infalls
adiabatically into the shell center. The total mass is now $M$. Using
the adiabatic invariant, it is known that the shell contracts to a new
radius, $\ra$, that is inversely proportional to the mass ratio
\citep{Blumenthal86}:
\begin{equation}
\label{eq:ac2}
  \frac{\ra}{\ri}=(1-\fin), \;\;\;  \frac{\Ma}{\Mi}=\frac{1}{(1-\fin)}.
\end{equation}

\subsubsection{Impulsive outflow}
Assume that gas of mass  $\fout M$ is removed instantaneously from
the centre of the sphere.  In the impulse approximation, the
instantaneous velocities remain as they were prior to the outflow:
\begin{equation}
  \Va^2 = G M / \ra,
\end{equation}
namely the system is temporarily out of virial equilibrium. The total
energy (potential plus kinetic) immediately after the gas removal is
thus
\begin{equation}
E = -\frac{G(1-\fout)^2 M^2}{\ra} + \frac{1}{2} (1-\fout)M\Va^2,
  \end{equation}
with $\Va$ given above.  Then the shell expands while conserving
energy to a new radius $\rf$ in a new virial equilibrium (total energy
= half the potential energy), so the energy is 
\begin{equation}
  E=-\frac{1}{2} \frac{G (1-\fout)^2 M^2}{\rf}.
\end{equation}
Equating the energies and inserting $\Va$ we obtain
\begin{equation}
  \label{eq:ie}
  \frac{\rf}{\ra} = \frac{1-\fout}{1-2\fout}, \;\;\; \frac{\Mf}{\Ma}=1-\fout.
\end{equation}
In the extreme case of losing half the mass, $\fout=1/2$, we get
$\rf \rightarrow \infty$, namely the system becomes unbound. In the limit
$\fout \ll 1$, to first order in $\fout$, $\rf/\ra \simeq 1+\fout$. So
to first order in $\fout$, the effect is the same as in an adiabatic
expansion, which would have been $\rf/\ra=(1-\fout)^{-1}\simeq (1 +
\fout)$.  However, the higher orders in $\fout$ make the impulse
expansion stronger than the adiabatic case.

One can show that the expansion for one episode with a given $\fout$ is
larger than the expansion for two episodes with $\fout/2$ each. And thus
the net outflow is not necessarily a robust predictor of the halo
response.

\subsubsection{Adiabatic contraction and impulsive expansion}

Now let us assume an adiabatic inflow with mass fraction $\fin$ followed
by an instantaneous outflow of mass fraction $\fout$, with the inflow
and outflow fractions related by
\begin{equation}
  \fout=\beta\fin.
\end{equation}
Combining equations (\ref{eq:ac}) \& (\ref{eq:ie}), we obtain
\begin{equation}
  \frac{\rf}{\ri}=\frac{(1-\beta\fin)(1-\fin)}{1-2\beta\fin}, \;\;\;
  \frac{\Mf}{\Mi}=\frac{1-\beta\fin}{1-\fin}.
\end{equation}

For equal inflow and outflow fractions $\fout=\fin\equiv f$ ($\beta=1$), the
result is net expansion. For net inflow ($\beta < 1$), for small
values of $f$ there is a net contraction, while for large values of
$f$ there is net expansion (See Fig.~\ref{fig:toymodel}). The boundary
occurs at $\fout=1-\beta$ and $\fin = \beta^{-1} -1$.  Since
$0 \le \fin < 1$, contraction occurs at all values of $\fin$ when
$\beta < 0.5$, i.e., the inflow fraction is more than double the
outflow fraction.

For $\fin \ll 1$ we get
\begin{equation}
\frac{\rf}{\ri} \simeq 1 - (1-\beta)\fin + \beta(2\beta -1) \fin^2,
\end{equation}
and thus for $\beta=1$ the effect is second order in $\fin$.

\subsubsection{Cycles of inflow and outflow}
If the removed gas remains bound and falls back to the centre, or if
fresh gas accretes, more than one cycle of inflow and outflow can
occur \citep[e.g.,][]{Read05}.
An upper limit to the number of cycles  is set by the dynamical
time. At the virial radius $t_{\rm dyn}\sim 1$ Gyr $h(z)^{-1}$, where
$h(z) = H(z)/100 \kmsmpc$, is the Hubble parameter.  Thus if the gas
reaches the virial radius, roughly 10 cycles are possible in a Hubble
time.  Since circular velocity is approximately constant inside a dark
matter halo, the dynamical time is smaller at smaller radii: $t_{\rm
  dyn}\sim 1 {\rm Gyr} (R/R_{200}) h(z)^{-1}$.  At the scale of the
half-mass radius of the stars $\sim 0.01 R_{200}$, the dynamical time
is thus $\sim 10$ Myr. Thus $\sim 1000$  cycles are in principle
possible in a Hubble time.

Fig.~11 of \citet{Tollet16} shows the mass fractions of stars gas and
dark matter inside 2kpc as a function of time for three NIHAO haloes
of mass $\sim 10^{10}, 10^{11}$, and $10^{12}\Msun$. At the time
resolution of the plot, we see $\sim 10$ cycles in a Hubble time where
the gas fraction varies by $\sim 0.1$.  

If each cycle has a different $\fin$ and $\fout$, then the net
contraction factor $\rf/\ri$, is simply the product of the contraction
factors of the separate episodes. A special case is when the same
cycle is repeated $N$ times, for which
\begin{equation}
  \frac{\rf}{\ri}=\left[\frac{(1-\fin)(1-\beta\fin)}{(1-2\beta\fin)}\right]^N,
  \;\;\;
\frac{\Mf}{\Mi}=\left[\frac{1-\beta\fin}{1-\fin}\right]^N.
\end{equation}
For $N$ large enough, the expansion (for $\fout > 1-\beta$) can be as
large as desired. For $\beta=1$ (i.e., equal inflow and outflow
fractions) and $f \ll 1$ it becomes
\begin{equation}
  \frac{\rf}{\ri} \simeq 1 + N f^2.
\end{equation}

For small values ($\fout\lta 0.1$), inflows and outflows have almost
symmetric effects, leading to no net change in the halo profile even
for large $N$.  For large values of $\fout$, the expansive effects of
outflows overcomes the contractive effects of inflows.  In the case of
equal inflow and outflow, the net effect is expansion (upper right
panel in Fig.~\ref{fig:toymodel}). However, unless $\fout$ is close to
0.5, or $N$ is large, the net change is only small. If the net effect
is inflow (lower middle panel), the halo will tend to contract for
small values of $\fout$ (corresponding to large radii) and expand for high
values of $\fout$ (corresponding to small radii). The example of
$\beta=0.83$ is shown.  In this case the maximum contraction occurs
for $\fout\sim 0.1$, while no change occurs at $\fout\sim 0.17$, and expansion
for higher values of $\fout$.

\begin{figure}
\centerline{
\psfig{figure=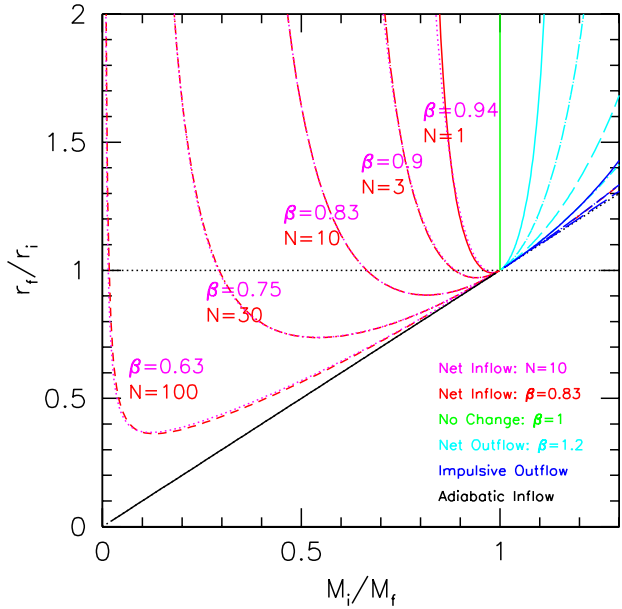,width=0.48\textwidth}}
\caption{Results in the halo response plane for a toy model of $N$
  cycles of adiabatic inflow followed by impulsive outflow. Colour and
  line types are the same as in Fig.~\ref{fig:toymodel}. Adiabatic
  contraction corresponds to the black diagonal line. For the case of
  net inflow very similar tracks can be found for different
  combinations of $\beta$ and $N$. Red lines show models with
  $\beta=0.83$ and $N$ varying from 1 to 100, while magenta lines show
  models with $N=10$ and $\beta$ varying from $0.94$ to $0.63$.}
  \label{fig:toymodel2}
\end{figure}

\subsubsection{Degeneracy between number of cycles and flow ratio }

Very similar tracks in the halo response mass radius plane can be
achieved with different combinations of $N$ and $\beta$ (see
Fig.~\ref{fig:toymodel2}). This degeneracy can be characterized by
comparing the mass ratios where there is no change in the radii:
$\rf/\ri=0$ when  $\fin = \beta^{-1} -1$, and $\Mi/\Mf = (2\beta
-1)^N/\beta^{2N}$. Thus if we wish to keep an integer number of
cycles, it makes sense to express $\beta$ in terms of $\Mi/\Mf$ and
$N$, thus $\beta = (1 - \sqrt{1 - (\Mi/\Mf)^{1/N}}) /
(\Mi/\Mf)^{1/N}$.  For example if we set $\Mi/\Mf=0.5$, we get the
degenerate solutions: $(N,\beta)=(1,0.586);(10,0.794); (100,0.923)$.
This degeneracy means we cannot uniquely infer the number of cycles
and inflow to outflow fraction based on the path of simulations in the
halo response plane.

\subsubsection{Bathtub model}

The bathtub model \citep{Dekel14} allows us to express $\beta$ for the
galaxy in terms of the star formation efficiency via
\begin{equation}
\beta = 1- \eSF/p
\end{equation}
where $p$ is the penetration factor: $p$=inflow rate on to galaxy /
inflow rate into halo.  For $\sim 10^{12}\Msun$ haloes at $z\sim 2$,
\citet{Dekel13} finds $p\sim 0.5$. For simplicity here we assume
$p=1$. In future, we will calibrate this as a function of mass and
redshift against our simulations.  The bathtub model can also be used
to link other properties of the outflow, such as the mass loading
factor, to the halo response (see Dekel \etal, in preparation).

\subsubsection{Comparison to simulations}

The halo response at a given radius, is thus determined by the value
of $\fout$ and $\fin$ at that radius and by $N$.  Consider the simple
case of the removal of an amount of gas from near the centre of the
halo to beyond the virial radius.  Since the mass removed is fixed,
while the total mass at each radius increases, $\fout$ decreases
monotonically with radius.  Thus, the response at different radii
within the halo can be spanned by varying $\fout$ from $\fout \ll 1$
at large radii to larger values up to $\fout = 1/2$ (where $\rf/\ri$
explodes in the impulse limit).

Fig.~\ref{fig:toymodel2} shows the models from Fig.~\ref{fig:toymodel}
in the halo response plane.  The red lines show the case of
$\fin=1.2\fout$, i.e., $\beta\simeq 0.83$, with the number of cycles
varying from $N=1$ to $100$.  The toy model has a similar qualitative
behaviour as the simulations.  We can understand why the tracks reach a
minimum $\rf/\ri$ and then increase as $\Mi/\Mf$ decreases: at smaller
radii (and hence smaller $\Mi/\Mf$) the local gas fraction ($ M_{\rm
  gas}(<r)/M_{\rm tot}(r)$) is larger, and thus the effects of
impulsive outflows are larger than the contractive effects of inflows.

\begin{figure}
\centerline{
\psfig{figure=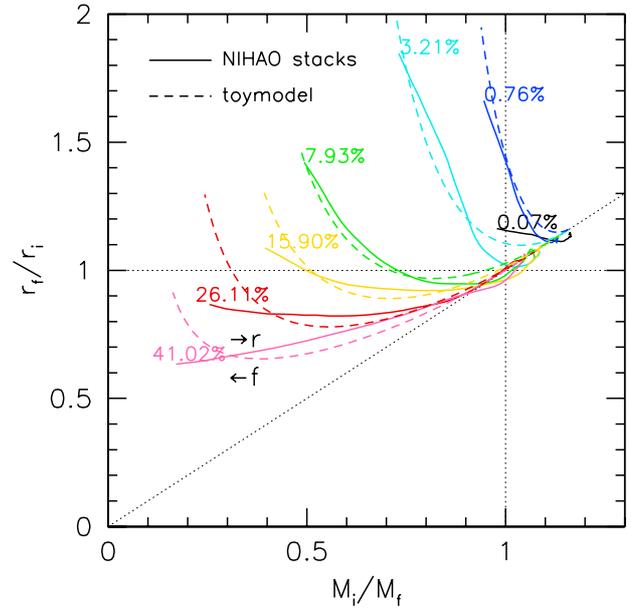,width=0.48\textwidth}}
\caption{ Average halo response for NIHAO haloes binned by star
  formation efficiency (solid lines) compared to the toy model (dashed
  lines), see the text for details. The arrows show the direction of
  increasing radius, $r$, in the simulation, and outflow fraction,
  $f$, in the toy model. }
  \label{fig:rrmm3}
\end{figure}

Fig.~\ref{fig:rrmm3} compares the average halo
response for galaxies with a given star formation efficiency with our
toy model. Here we use the bathtub model to set $\beta=1-\eSF$ and
vary the number of inflow and outflow cycles to roughly match the
simulations.  For $\eSF=(0.410, 0.261, 0.159, 0.079, 0.032, 0.008)$ we
find $N=(3, 10, 25, 90, 200, 1000)$.  Using the bathtub model we can
infer the mass-loading factors using $\eta=0.5\beta/(1-\beta)$,
obtaining $\eta=(0.72, 1.42, 2.64, 5.83, 15.1, 62)$.  Additionally, in
order to match the simulations at $\Mi/\Mf>1$, we add an episode of
adiabatic expansion, with $f_{\rm out}=0.17(1-2\eSF)$, corresponding
to permanent gas loss from the system.

In detail the expansive effects are too strong at small $\Mi/\Mi$
(small radii).  This may be due to break down in the impulse
approximation for high gas fractions,  or if the number of cycles,
$N$, or the outflow-to-inflow ratio, $\beta$, varies with radius, or if 
$f$ or $\beta$ vary time stochastically or systematically with time.

\subsubsection{Comparison to Pontzen \& Governato (2012)}

\citet{Pontzen12} present a model for the response of the
dark matter halo to a fluctuating potential sourced by supernova
driven gas flows.  Assuming a spherical potential
of the form $V(r,t) = V_0(t) r^{-1}$, their equation 6 gives the energy gain
in terms of the potential change
\begin{equation}
  \frac{\Delta E}{E_0} = -2\left(\frac{\Delta
    V_0}{V_0}\right)^2 = -2 f^2.
  \label{eq:PG12}
\end{equation}
Where the last equality is for the isolated shell case we study which assumes 
$V_0\propto GM/r$, thus $\Delta V / V = \Delta M / M = f$ at fixed radius.

\begin{figure*}
  \centerline{
    \psfig{figure=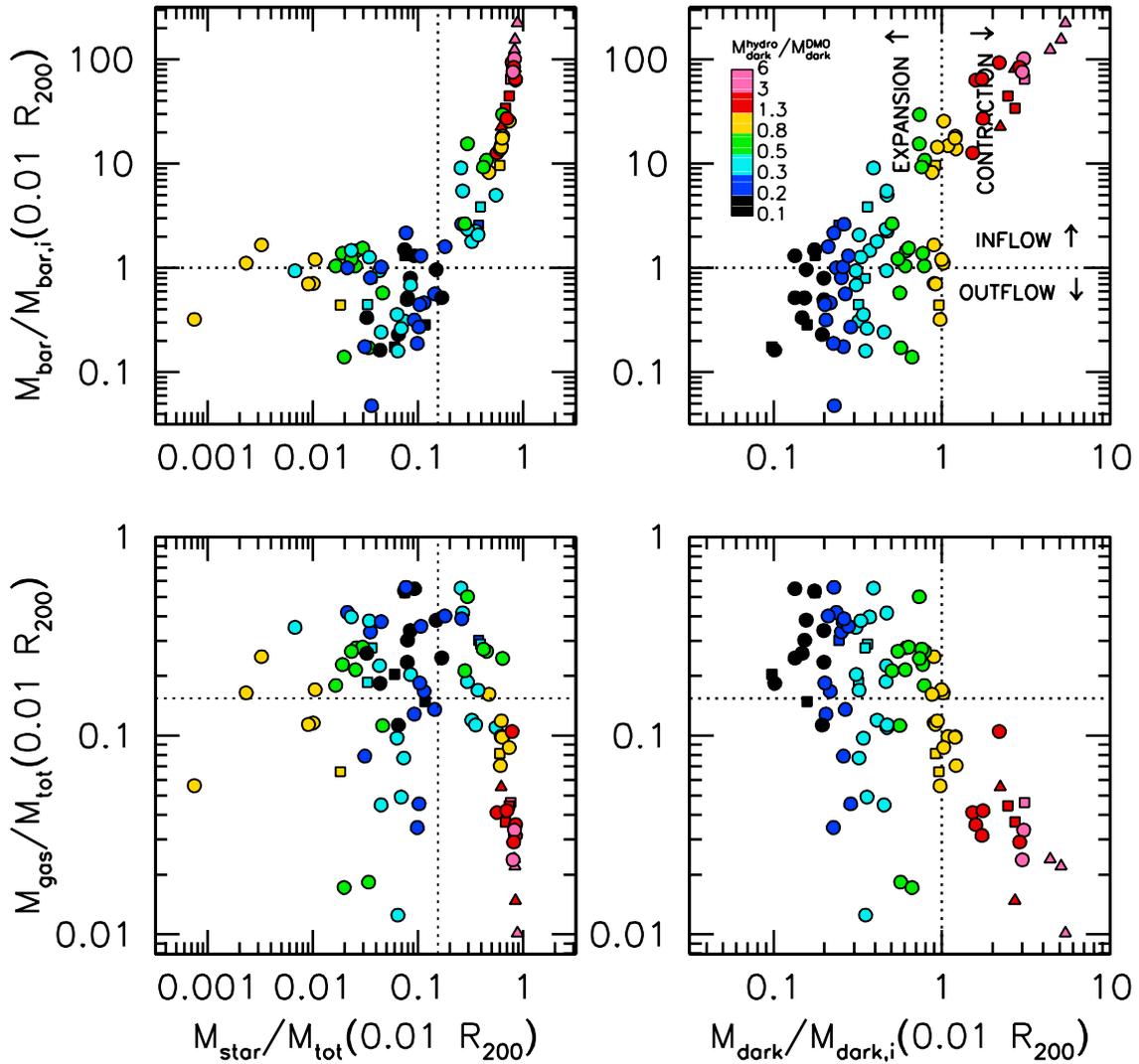,width=0.85\textwidth} }
  \caption{Correlations between halo response, star formation
    efficiency, gas fraction, and baryonic mass ratio, all measured at
    1\% of the virial radius.   Points are colour coded by the halo
    response. Dotted lines correspond to either unity or the cosmic
    baryon fraction ($\fbar=0.154$).}
  \label{fig:vv4}
\end{figure*}

We derive a similar result in the context of our model under similar
assumptions. By comparing the energy derived from the virial theorem
before and after the episode of adiabatic inflow and impulsive outflow
gives: $\Ef = \Ei (\ri/\rf)$, and thus the change in energy is $\Delta
E = \Ei (\ri/\rf -1)$. Assuming small $f$ and $\beta=1$  gives
$\Delta E / \Ei = -f^2$. Re-deriving the equation for impulsive
inflows and outflows, we find $\Delta E / \Ei = -2f^2$, in agreement
with equation (\ref{eq:PG12}). The overall effect when using impulsive
inflows is larger because they have less of a contractive effect than
adiabatic inflows.


\section{Discussion}
\label{sec:discussion}

We now discuss the various phases of halo response and how they are
related to physical properties of the galaxies. We then compare our
results with those found in the literature.

\subsection{Inflow versus outflow}

The upper panels of Fig.~\ref{fig:vv4} show ratio between baryonic
mass inside $0.01 R_{200}$ in the hydro and DMO simulations versus
stellar mass fraction within $0.01 R_{200}$ (upper left) and change in
dark matter masses within $0.01 R_{200}$ (upper right).  Points are
colour coded by change in dark matter mass. No change is yellow, black
is the most expansion and magenta is the most contraction.  The
upper-right panel is divided into four quadrants. Lower right is
unphysical (contracted dark matter with expanded baryons) and
reassuringly all simulations avoid this region. Lower left is
expansion in both dark matter and baryons. The upper left is when the
dark matter expands but there is net inflow of baryons, with greater
inflow resulting in less expansion. The trend continues to the upper
right where more inflow leads to more contraction of the dark
matter. Contraction starts when the baryonic mass increases by a
factor of $~\sim 10$. At this point, the baryons start to dominate the
total mass budget at small radii. 

The upper-left panel shows that at low \eSF there is no net change in
baryon content.  For intermediate $\eSF \sim 0.01$, there is actually
less baryons inside $0.01R_{200}$ than if the baryons trace the DMO
simulation.  Thus, a significant fraction of the baryons
that cooled have been ejected. Perhaps, not surprisingly, these
galaxies have the strongest expansion. Above $\eSF = 0.03$ we see net
inflow of baryons, and the stars start to dominate the baryon budget
so there is a tight correlation between the axes.

\subsection{The role of gas}
         
Gas plays an important role in two ways: (1) gas accretion drives
contraction; and (2) gas outflows cause significant expansion only if
the gas fraction is high enough to allow $\fout\gta 0.1$ (see
Fig.\ref{fig:toymodel}).  The lower panels of Fig.~\ref{fig:vv4} shows
the gas fraction versus stellar mass fraction (left) and the dark matter
mass ratio (right). All quantities are measured at 1\% of the virial
radius.  Points are colour coded by the dark matter mass ratio.  Since
the gas fraction can fluctuate wildly on small time-scales due to SN
feedback \citep[e.g.,][]{Tollet16}, we focus on the average gas
fractions, which correlate with both star formation efficiency and the
halo response. We verified that the galaxies with halo expansion and
low gas fractions inside of $0.01R_{200}$ are in fact gas rich, with
global cold gas to stellar mass ratios in excess of 1. 
Contraction only occurs when the gas fraction  is low ($\lta 0.1$),
while the most expansion occurs when the gas fraction is high $\sim
0.3$. These trends are consistent with the expectation of our toy
model.

\subsection{The role of star formation}

Star formation play two roles: (1) star formation drives expansion from
the high mass stars that turn into supernovae. (2) However, since stars
are collisionless, once formed they behave the same way as the dark
matter and are effectively added to the overall potential. Subsequent
star formation episodes will face a deeper potential, and find it
harder to escape. If there is no new supply of gas, the reduction of
gas will also weaken the expansive effects of any outflows.

\subsection{Comparison to previous results}

Several authors have studied the halo response in cosmological galaxy
formation simulations finding contraction in haloes of mass $\sim
10^{12}\Msun$ \citep{Gnedin04, Abadi10, Pedrosa10}.  It should be
noted that these studies suffered from over-cooling and small sample
sizes (the \citealt{Abadi10} simulations do not include star
formation). So it is not clear how relevant they are to typical
galaxies.  Nevertheless, in our simulations, when the star formation
efficiency is high we also find haloes contract strongly, in broad agreement
with these earlier studies.

\citet{DiCintio14a}, used  the MaGICC simulations \citep{Stinson13} to
study the dark matter density slopes in the inner 1-2 per cent of the virial
radius.  They found a non-linear dependence between the inner slope
and stellar to halo mass.  This result has been confirmed with a
larger sample of galaxies from NIHAO \citep{Tollet16}, and a small
sample of higher resolution galaxies from FIRE \citep{Chan15}. Since
we use the same simulations  our results are consistent with
\citet{DiCintio14a} and \citet{Tollet16}.  Going a step further,
\citet[][DC14]{DiCintio14b} present a dark matter density profile
dependent on $\eSF$ based on 10 simulations with the fiducial MaGICC
feedback.  As discussed in \S\ref{sec:implications}, the DC14 model
likely overestimates the expansion at $\eSF\sim0.01$ and overestimates
the contraction at $\eSF\gta 0.2$, although the qualitative dependence
of halo contraction and expansion on $\eSF$ remains the same.

The EAGLE project \citep{Schaye15} has formed galaxies with reasonable
star formation efficiencies and galaxy sizes over a wide range of
masses. When looking at average halo profiles in bins of halo mass
from $M_{200}=10^{10}$ to $10^{14}\Msun$, \citet{Schaller15} finds
almost no difference in dark matter profiles between hydro and DMO
simulations. As touched on in \S~\ref{sec:haloresponse}, an important
issue is resolution. For haloes of mass $\Mhalo \sim 10^{11}\Msun$ (the
scale where we find the most expansion in our simulations), their
convergence radius is 5\% of the virial radius. Looking at
Fig.~\ref{fig:vv2} we find very little expansion (or contraction)
above this scale, and thus their lack of expansion is expected. At the
MW mass scale ($\Mhalo \sim 10^{12}\Msun$), their simulations
find very little change in the halo mass profile, as we do.  So our
respective simulations may actually be more consistent than at first
apparent. 

Another potential source of differences in halo response is the
threshold for star formation. NIHAO and MaGICC uses $n_{\rm th}\simeq
10$ while EAGLE adopts a much lower threshold of $n_{\rm th}\lta 0.1$.
A lower threshold is expected to result in a more uniform star
formation history, with fewer starbursts, and hence fewer outflow
events with high $\fout$, and thus weaker expansion.  However, Dutton
\etal (2015) found halo expansion that follows the NIHAO results in
the progenitors of massive elliptical galaxies using a threshold of
$n_{\rm th}\sim 1$ cm$^{-3}$, and the FIRE simulations \citep{Chan15}
are also consistent with NIHAO (from dwarfs to MW mass haloes) and use
a higher threshold of $n_{\rm th} > 10 - 100$ cm$^{-3}$. Thus the halo
response does not seem strongly sensitive to the star formation
threshold, at least for $n_{\rm th} > 1$ cm$^{-3}$.  Determining how
sensitive the halo response is to the sub-grid model for star
formation and feedback needs to be studied in greater detail.


\section{Summary}
\label{sec:summary}

We use $\sim 100$ hydrodynamical cosmological zoom-in simulations from
NIHAO, MaGICC, and MUGS \citep{Wang15, Stinson13, Stinson10} to
investigate the response of CDM haloes to baryonic galaxy
formation processes. We define the halo response as the change in dark
matter mass profiles, at a given radius, between hydrodynamical and
dissipationless simulations.  We summarize our results as follows:

\begin{itemize}
\item There is a much wider diversity of dark matter halo profiles in
  hydrodynamical simulations than dissipationless simulations
  (Fig.~\ref{fig:vv}). The halo response to galaxy formation thus breaks
  the scale free nature of CDM haloes.
  
\item Dark matter haloes are only significantly modified at radii
  below $\sim 0.05R_{200}$ (Fig.~\ref{fig:vv2}), and thus sufficiently
  high resolution is required to see the effects.

\item The halo response is correlated with halo mass, $\Mhalo$,
  stellar mass, $\Mstar$,    integrated star formation efficiency,
  $\eSF=(\Mstar/\Mhalo)/(\Omegab/\Omegam)$, and the compactness of the
  galaxy, $\eR=\rhalf/R_{200}$. The correlation with $\eSF$ has the
  smallest scatter (0.15 dex), while the correlation with galaxy
  compactness is well described with a power-law above $\eSF\sim
  10^{-3}$ (Fig.~\ref{fig:vv3}).
  
\item The halo response at any radius is proportional to
  $\eR/\eSF^{0.13}$, with a shallower slope at larger radii
  (Fig.~\ref{fig:fp}). We provide an empirical formula
  (Fig.~\ref{fig:fp2}), section 3.4, that reproduces the response of
  the dark matter haloes on scales from $0.01 R_{200}$ to $R_{200}$
  (Fig.~\ref{fig:xmm2}).

\item The path of galaxies in the $\rf/\ri$ versus $\Mi/\Mf$ plane of the
  adiabatic contraction formalism depends on $\eSF$
  (Fig.~\ref{fig:rrmm}). Galaxies with high star formation
  efficiencies ($\eSF\gta 0.5$) have contraction close to that of the
  adiabatic contraction formalism of \citet{Gnedin04}. For lower
  $\eSF$ the contraction is progressively weaker, first leading to
  expansion at small $\Mi/\Mf$, and then at all $\Mi/\Mf$. Finally
  when $\eSF$ is very low, the halo response is weak.
  
\item The paths of galaxies in the $\rf/\ri$ versus $\Mi/\Mf$ plane can be
  understood by a toy model consisting of several cycles of adiabatic
  inflows followed by impulsive outflows (Fig.~\ref{fig:rrmm3}). Pure
  inflows result in contraction, while pure outflows result in
  expansion. For equal inflow and outflow fractions, $f$,  the net
  effect is expansion. For net inflow, contraction occurs for small
  $f$ (i.e., large radii), while expansion occurs for large $f$ (i.e.,
  small radii). 

\item For significant halo expansion to occur the local gas fraction
  needs to be $\gta 0.1$ (Fig.~\ref{fig:vv4}).  Even if there is
  sufficient energy to drive an outflow, if there is not enough gas,
  the expansive effects of the outflow will be minimal.
  
\end{itemize}

Our results on the halo response to galaxy formation can be coupled to
results from dissipationless simulations to make predictions for the
structure of CDM haloes. Since our model for star formation and
feedback is by no means unique, at present it is not clear whether our
fitting formula for the halo response is applicable to other subgrid
models. Thus, our predictions are not yet at the stage that they can be
used to falsify the CDM model.  Nevertheless, the regularities present
in our simulations suggest that, despite the many complex processes
involved in galaxy formation, a fully predictive model for the
structure of CDM haloes may soon be possible. 

Determining on which mass and radial scales different subgrid models
agree / disagree on the halo response is an important subject for
future studies. Also it would be interesting to find additional
observable/modelable parameters, besides $\eSF$ and $\eR$, that the
halo response correlates with.

\section*{Acknowledgements} 
This research was carried out on the High Performance Computing
resources at New York University Abu Dhabi; on the {\sc theo} cluster
of the Max-Planck-Institut f\"ur Astronomie and the {\sc hydra}
cluster at the Rechenzentrum in Garching; and the MW
supercomputer, which is funded by the Deutsche Forschungsgemeinschaft
(DFG) through Collaborative Research Center (SFB 881) `The Milky Way
System' (subproject Z2) and  hosted and co-funded by the J\"ulich
Supercomputing Center (JSC). We greatly appreciate the contributions
of all these computing allocations.
This work was supported by Sonderforschungsbereich SFB 881 `The Milky
Way System' (subprojects A1 and A2) of the German Research Foundation
(DFG).
The analysis made use of the {\sc pynbody} package \citep{Pontzen13}.
AD acknowledges support by ISF grant 24/12, by the I-CORE Program of
the PBC - ISF grant 1829/12, by BSF grant 2014-273, and by NSF grants
AST-1010033 and AST-1405962.
XK acknowledge the support from NSFC project no.11333008 and the
`Strategic Priority Research Program the Emergence of Cosmological
Structures' of the CAS(No.XD09010000).


\label{lastpage}

\begin{thebibliography}{}


\bibitem[Abadi et al.(2010)]{Abadi10} Abadi, M.~G., Navarro, 
J.~F., Fardal, M., Babul, A., \& Steinmetz, M.\ 2010, \mnras, 847 


\bibitem[Baldry et al.(2012)]{Baldry12} Baldry, I.~K., Driver, 
S.~P., Loveday, J., et al.\ 2012, \mnras, 421, 621 

\bibitem[Behroozi et al.(2013)]{Behroozi13} Behroozi, P.~S.,
  Wechsler, R.~H., \& Conroy, C.\ 2013, \apj, 770, 57

\bibitem[Blumenthal et al.(1986)]{Blumenthal86} Blumenthal,
  G.~R., Faber, S.~M., Flores, R., \& Primack, J.~R., 1986, ApJ, 301, 27

\bibitem[Bovy \& Rix(2013)]{BovyRix13} Bovy, J., \& Rix, H.-W.\ 2013, \apj, 779, 115 

\bibitem[Brook et al.(2011)]{Brook11} Brook, C.~B., Governato, 
F., Ro{\v s}kar, R., et al.\ 2011, \mnras, 415, 1051 

\bibitem[Brook et al.(2012)]{Brook12} Brook, C.~B., Stinson, G., Gibson, B.~K., Wadsley, J., \& Quinn, T.\ 2012, \mnras, 424, 1275 

\bibitem[Brook et al.(2016)]{Brook16} Brook, C.~B., Santos-Santos, I., \& Stinson, G.\ 2016, \mnras, 459, 638 
  
\bibitem[Brooks \& Christensen(2016)]{Brooks16} Brooks, A., \& Christensen, C.\ 2016, Galactic Bulges, 418, 317 

\bibitem[Bullock  et al.(2001)]{Bullock01}  Bullock, J.~S.,
  Kolatt,  T.~S.,  Sigad,  Y.,  Somerville,  R.~S.,  Kravtsov,  A.~V.,
  Klypin, A.~A., Primack, J.~R., \&  Dekel, A.\ 2001, MNRAS, 321, 559

 \bibitem[Butsky et al.(2015)]{Butsky15} Butsky, I., Macci{\`o}, A.~V., Dutton, A.~A., et al.\ 2015, arXiv:1503.04814 
 

\bibitem[Chabrier(2003)]{Chabrier03} Chabrier, G.\ 2003, PASP, 
115, 763 

 \bibitem[Chan et al.(2015)]{Chan15} Chan, T.~K., Kere{\v s}, 
D., O{\~n}orbe, J., et al.\ 2015, \mnras, 454, 2981 
 
\bibitem[Cole et al.(2011)]{Cole11} Cole, D.~R., Dehnen, W., 
\& Wilkinson, M.~I.\ 2011, \mnras, 416, 1118 

\bibitem[Conroy \& Wechsler(2009)]{Conroy09} Conroy, C., \&
  Wechsler, R.~H.\ 2009, \apj, 696, 620

\bibitem[Courteau \& Dutton(2015)]{Courteau15} Courteau, S., \&
  Dutton, A.~A.\ 2015, \apjl, 801, L20 

  \bibitem[Crain et al.(2015)]{Crain15} Crain, R.~A., Schaye, J., 
Bower, R.~G., et al.\ 2015, \mnras, 450, 1937 


\bibitem[Croton et al.(2006)]{Croton06} Croton, D.~J., Springel, 
  V., White, S.~D.~M., et al.\ 2006, \mnras, 365, 11

  

\bibitem[de Blok et al.(2001)]{deBlok01} de Blok, W.~J.~G., 
  McGaugh, S.~S., Bosma, A., \& Rubin, V.~C.\ 2001, \apjl, 552, L23

\bibitem[Dekel \& Silk(1986)]{Dekel86} Dekel, A., \& Silk, J.\ 1986, \apj, 303, 39 

  
 \bibitem[Dekel et al.(2003)]{Dekel03} Dekel, A., Devor, J., 
\& Hetzroni, G.\ 2003, \mnras, 341, 326 

\bibitem[Dekel et al.(2013)]{Dekel13} Dekel, A., Zolotov, A., Tweed, D., et al.\ 2013, \mnras, 435, 999 

\bibitem[Dekel \& Mandelker(2014)]{Dekel14} Dekel, A., \& Mandelker, N.\ 2014, \mnras, 444, 2071 

\bibitem[Diemer  \& Kravtsov(2015)]{Diemer15} Diemer, B.,
  \& Kravtsov, A.~V.\ 2015, \apj, 799, 108 

  \bibitem[Di Cintio et al.(2014a)]{DiCintio14a} Di Cintio, A., Brook, 
C.~B., Macci{\`o}, A.~V., et al.\ 2014a, \mnras, 437, 415 

\bibitem[Di Cintio et al.(2014b)]{DiCintio14b} Di Cintio, A., Brook, 
C.~B., Dutton, A.~A., Macci\`o, A.~V., Stinson, G.~S., Knebe, A.\ 2014b, \mnras, 441, 2986
 
\bibitem[Duffy et al.(2010)]{Duffy10} Duffy, A.~R.,
  Schaye, J., Kay, S.~T., Dalla Vecchia, C., Battye, R.~A., \& Booth,
  C.~M.\ 2010, \mnras, 405, 2161

\bibitem[Dutton et al.(2007)]{Dutton07} Dutton, A.~A., van den 
Bosch, F.~C., Dekel, A., \& Courteau, S.\ 2007, \apj, 654, 27 

 \bibitem[Dutton(2009)]{Dutton09} Dutton, A.~A.\ 2009, \mnras, 
396, 121 

\bibitem[Dutton et al.(2013a)]{Dutton13a} Dutton, A.~A., Treu, T., Brewer, B.~J., et al.\ 2013, \mnras, 428, 3183 
 
\bibitem[Dutton et al.(2013b)]{Dutton13b} Dutton, A.~A., Macci{\`o}, A.~V., Mendel, J.~T., \& Simard, L.\ 2013, \mnras, 432, 2496 

\bibitem[Dutton \& Treu(2014)]{Dutton14a} Dutton, A.~A., \& Treu, T.\ 2014, \mnras, 438, 3594 

\bibitem[Dutton \& Macci{\`o}(2014)]{Dutton14b} Dutton,
  A.~A., \& Macci{\`o}, A.~V.\ 2014, \mnras, 441, 3359

\bibitem[Dutton et al.(2015)]{Dutton15} Dutton, A.~A., 
Macci{\`o}, A.~V., Stinson, G.~S., et al.\ 2015, \mnras, 453, 2447 

\bibitem[Dutton et al.(2016)]{Dutton16} Dutton, A.~A., Macci{\`o}, A.~V., Frings, J., et al.\ 2016, \mnras, 457, L74 

  

\bibitem[Einasto(1965)]{Einasto65} Einasto, J.\ 1965, Trudy 
Astrofizicheskogo Instituta Alma-Ata, 5, 87 

\bibitem[El-Zant et al.(2001)]{El-Zant01} El-Zant, A.,
  Shlosman, I., \& Hoffman, Y.\ 2001, \apj, 560, 636

\bibitem[El-Zant et al.(2016)]{El-Zant16} El-Zant, A., Freundlich, J., \& Combes, F.\ 2016, arXiv:1603.00526 




\bibitem[Garrison-Kimmel et al.(2016)]{Garrison-Kimmel16} Garrison-Kimmel, S., Bullock, J.~S., Boylan-Kolchin, M., \& Bardwell, E.\ 2016, arXiv:1603.04855 
  
\bibitem[Gill et al.(2004)]{Gill04} Gill, S.~P.~D.,
  Knebe, A., \& Gibson, B.~K.\ 2004, \mnras, 351, 399
  
\bibitem[Gnedin  et   al.(2004)]{Gnedin04}  Gnedin,  O.~Y.,
Kravtsov, A.~V., Klypin, A.~A., \& Nagai, D.\ 2004, \apj, 616, 16

\bibitem[Goerdt et al.(2006)]{Goerdt06} Goerdt, T., Moore, B., 
Read, J.~I., Stadel, J., \& Zemp, M.\ 2006, \mnras, 368, 1073 

\bibitem[Governato et al.(2012)]{Governato12} Governato, F., 
Zolotov, A., Pontzen, A., et al.\ 2012, \mnras, 422, 1231 

\bibitem[Gustafsson et al.(2006)]{Gustafsson06} Gustafsson, M., 
Fairbairn, M., \& Sommer-Larsen, J.\ 2006, \prd, 74, 123522 

\bibitem[Gutcke et al.(2016)]{Gutcke16} Gutcke, T.~A., Stinson, G.~S., Macci{\`o}, A.~V., Wang, L., \& Dutton, A.~A.\ 2016, arXiv:1602.06956 

  

\bibitem[Hammer et al.(2007)]{Hammer07} Hammer, F., Puech, M., Chemin, L., Flores, H., \& Lehnert, M.~D.\ 2007, \apj, 662, 322 

\bibitem[Hopkins et al.(2014)]{Hopkins14} Hopkins, P.~F., Kere{\v 
s}, D., O{\~n}orbe, J., et al.\ 2014, \mnras, 445, 581 

\bibitem[Hudson et al.(2015)]{Hudson15} Hudson, M.~J., Gillis, 
B.~R., Coupon, J., et al.\ 2015, \mnras, 447, 298 

  

\bibitem[Jardel \& Sellwood(2009)]{Jardel09} Jardel, J.~R.,
  \& Sellwood, J.~A.\ 2009, \apj, 691, 1300

\bibitem[Johansson et al.(2009)]{Johansson09} Johansson,
  P.~H., Naab, T., \& Ostriker, J.~P.\ 2009, \apjl, 697, L38



\bibitem[Keller et al.(2014)]{Keller14} Keller, B.~W., Wadsley, 
  J., Benincasa, S.~M., \& Couchman, H.~M.~P.\ 2014, \mnras, 442, 3013

\bibitem[Klypin et al.(2011)]{Klypin11} Klypin, A.~A., 
Trujillo-Gomez, S., \& Primack, J.\ 2011, \apj, 740, 102 

\bibitem[Knollmann \& Knebe(2009)]{Knollmann09} Knollmann,
  S.~R., \& Knebe, A.\ 2009, \apjs, 182, 608

  \bibitem[Kravtsov(2013)]{Kravtsov13} Kravtsov, A.~V.\ 2013, \apjl, 
764, L31 

  \bibitem[Kroupa et al.(1993)]{Kroupa93} Kroupa, P., Tout, C.~A., 
\& Gilmore, G.\ 1993, \mnras, 262, 545 


\bibitem[Leauthaud et al.(2012)]{Leauthaud12} Leauthaud, A., 
Tinker, J., Bundy, K., et al.\ 2012, \apj, 744, 159 

\bibitem[Ludlow et al.(2016)]{Ludlow16} Ludlow, A.~D., Bose, S., Angulo, R.~E., et al.\ 2016, \mnras, 460, 1214 



\bibitem[Macci{\`o} et al.(2008)]{Maccio08} Macci{\`o}, A.~V., 
Dutton, A.~A., \& van den Bosch, F.~C.\ 2008, \mnras, 391, 1940 

\bibitem[Macci{\`o} et al.(2012)]{Maccio12} Macci{\`o},
  A.~V., Stinson, G., Brook, C.~B., et al.\ 2012, \apjl, 744, L9

\bibitem[Macci{\`o} et al.(2015)]{Maccio15} Macci{\`o}, A.~V., 
Mainini, R., Penzo, C., \& Bonometto, S.~A.\ 2015, \mnras, 453, 1371 

\bibitem[Maller \& Dekel(2002)]{Maller02} Maller, A.~H., \& Dekel, A.\ 2002, \mnras, 335, 487 

\bibitem[Marinacci et al.(2014)]{Marinacci14} Marinacci, F., Pakmor, R., \& Springel, V.\ 2014, \mnras, 437, 1750 

  \bibitem[Martizzi et al.(2012)]{Martizzi12} Martizzi, D., 
Teyssier, R., Moore, B., \& Wentz, T.\ 2012, \mnras, 422, 3081 

\bibitem[Martizzi et al.(2013)]{Martizzi13} Martizzi, D., 
  Teyssier, R., \& Moore, B.\ 2013, \mnras, 432, 1947
  
\bibitem[Mashchenko et al.(2006)]{Mashchenko06} Mashchenko, S., 
Couchman, H.~M.~P., \& Wadsley, J.\ 2006, \nat, 442, 539 

\bibitem[Mashchenko et al.(2008)]{Mashchenko08} Mashchenko, S.,
  Wadsley, J., \& Couchman, H.~M.~P.\ 2008, Science, 319, 174

\bibitem[Moster et al.(2010)]{Moster10} Moster, B.~P., 
Somerville, R.~S., Maulbetsch, C., et al.\ 2010, \apj, 710, 903 

\bibitem[More et al.(2011)]{More11} More, S., van den Bosch, 
F.~C., Cacciato, M., Skibba, R., Mo, H.~J., \& Yang, X.\ 2011, \mnras, 410, 210 


\bibitem[Navarro et al.(1996)]{Navarro96} Navarro, J.~F., Eke, 
V.~R., \& Frenk, C.~S.\ 1996, \mnras, 283, L72 

\bibitem[Navarro  et  al.(1997)]{NFW97} Navarro,  J.~F.,
Frenk, C.~S., \& White, S.~D.~M.\ 1997, ApJ, 490, 493 (NFW)

\bibitem[Navarro et al.(2010)]{Navarro10} Navarro, J.~F., et al.\ 
2010, \mnras, 402, 21 

 \bibitem[Newman et al.(2013)]{Newman13} Newman, A.~B., Treu, T., 
Ellis, R.~S., \& Sand, D.~J.\ 2013, \apj, 765, 25 
 

\bibitem[Obreja et al.(2014)]{Obreja14} Obreja, A., Brook, C.~B., Stinson, G., et al.\ 2014, \mnras, 442, 1794 

\bibitem[Obreja et al.(2016)]{Obreja16} Obreja, A., Stinson, G.~S., Dutton, A.~A., et al.\ 2016, \mnras, 459, 467 
  
\bibitem[Oh et al.(2011)]{Oh11} Oh, S.-H., de Blok,
  W.~J.~G., Brinks, E., Walter, F., \& Kennicutt, R.~C., Jr.\ 2011,
  \aj, 141, 193


\bibitem[Pagels 
\& Primack(1982)]{Pagels82} Pagels, H., \& Primack, J.~R.\ 1982, Physical Review Letters, 48, 223 

\bibitem[Pedrosa et al.(2010)]{Pedrosa10} Pedrosa, S.,
  Tissera, P.~B., \& Scannapieco, C.\ 2010, \mnras, 402, 776

\bibitem[Pe{\~n}arrubia et al.(2016)]{Penarrubia16} Pe{\~n}arrubia, J., G{\'o}mez, F.~A., Besla, G., Erkal, D., \& Ma, Y.-Z.\ 2016, \mnras, 456, L54 
  
\bibitem[Planck Collaboration et al.(2014)]{Planck14} Planck Collaboration, Ade, P.~A.~R., Aghanim, N., et al.\ 2014, \aap, 571, A16 
  
\bibitem[Pontzen \& Governato(2012)]{Pontzen12} Pontzen, A., \& Governato, F.\ 2012, \mnras, 421, 3464 

\bibitem[Pontzen et al.(2013)]{Pontzen13} Pontzen, A., Ro{\v 
s}kar, R., Stinson, G., \& Woods, R.\ 2013, Astrophysics Source Code Library, 1305.002 

\bibitem[Power et al.(2003)]{Power03} Power, C., Navarro, 
J.~F., Jenkins, A., et al.\ 2003, \mnras, 338, 14 

 

\bibitem[Read \& Gilmore(2005)]{Read05} Read, J.~I., \&
Gilmore, G.\ 2005, \mnras, 356, 107

\bibitem[Reddick et al.(2013)]{Reddick13} Reddick, R.~M., 
Wechsler, R.~H., Tinker, J.~L., \& Behroozi, P.~S.\ 2013, \apj, 771, 30 

\bibitem[Romano-D{\'{\i}}az et al.(2008)]{Romano-Diaz08} 
Romano-D{\'{\i}}az, E., Shlosman, I., Hoffman, Y., 
\& Heller, C.\ 2008, \apjl, 685, L105 




\bibitem[Sawala et al.(2016)]{Sawala16} Sawala, T., Frenk, C.~S., Fattahi, A., et al.\ 2016, \mnras, 457, 1931 

\bibitem[Schaller et al.(2015)]{Schaller15} Schaller, M., Frenk, 
  C.~S., Bower, R.~G., et al.\ 2015, \mnras, 451, 1247

\bibitem[Schaye et al.(2015)]{Schaye15} Schaye, J., Crain,
  R.~A., Bower, R.~G., et al.\ 2015, \mnras, 446, 521

\bibitem[Sellwood(2008)]{Sellwood08} Sellwood, J.~A.\ 2008,
  \apj, 679, 379

\bibitem[Shen et al.(2010)]{Shen10} Shen, S., Wadsley, J., 
\& Stinson, G.\ 2010, \mnras, 407, 1581 

 \bibitem[Somerville et al.(2008)]{Somerville08} Somerville, R.~S.,
   Hopkins, P.~F., Cox, T.~J., Robertson, B.~E.,  \& Hernquist,
   L.\ 2008, \mnras, 391, 481 

\bibitem[Sonnenfeld et al.(2015)]{Sonnenfeld15} Sonnenfeld, A., Treu, T., Marshall, P.~J., et al.\ 2015, \apj, 800, 94 
  
\bibitem[Spergel 
\& Steinhardt(2000)]{Spergel00} Spergel, D.~N., \& Steinhardt, P.~J.\ 2000, Physical Review Letters, 84, 3760 

\bibitem[Spergel et al.(2007)]{Spergel07} Spergel, D.~N., Bean, 
R., Dor{\'e}, O., et al.\ 2007, \apjs, 170, 377 


\bibitem[Stadel et al.(2009)]{Stadel09} Stadel, J., Potter, D., 
Moore, B., et al.\ 2009, \mnras, 398, L21 

\bibitem[Stinson et al.(2006)]{Stinson06} Stinson, G., Seth, A., 
Katz, N., et al.\ 2006, \mnras, 373, 1074 

\bibitem[Stinson et al.(2010)]{Stinson10} Stinson, G.~S., Bailin, 
J., Couchman, H., et al.\ 2010, \mnras, 408, 812 

\bibitem[Stinson et al.(2013)]{Stinson13} Stinson, G.~S., Brook, 
C., Macci{\`o}, A.~V., et al.\ 2013, \mnras, 428, 129 

\bibitem[Stinson et al.(2015)]{Stinson15} Stinson, G.~S., Dutton, A.~A., Wang, L., et al.\ 2015, \mnras, 454, 1105 

  
\bibitem[Swaters et al.(2003)]{Swaters03} Swaters, R.~A., Madore, 
B.~F., van den Bosch, F.~C., \& Balcells, M.\ 2003, \apj, 583, 732 




\bibitem[Tissera et al.(2010)]{Tissera10} Tissera, P.~B., White, 
S.~D.~M., Pedrosa, S., \& Scannapieco, C.\ 2010, \mnras, 406, 922 

\bibitem[Tollet et al.(2016)]{Tollet16} Tollet, E., Macci{\`o}, A.~V., Dutton, A.~A., et al.\ 2016, \mnras, 456, 3542 

\bibitem[Trujillo-Gomez et al.(2015)]{Trujillo-Gomez15} Trujillo-Gomez, 
S., Klypin, A., Col{\'{\i}}n, P., et al.\ 2015, \mnras, 446, 1140 



\bibitem[Vogelsberger et al.(2014)]{Vogelsberger14} Vogelsberger, M., Genel, S., Springel, V., et al.\ 2014, \mnras, 444, 1518 

  

\bibitem[Wadsley et al.(2004)]{Wadsley04} Wadsley, J.~W., Stadel, 
J., \& Quinn, T.\ 2004, \na, 9, 137 

\bibitem[Walker \& Pe{\~n}arrubia(2011)]{Walker11} Walker,
  M.~G., \& Pe{\~n}arrubia, J.\ 2011, \apj, 742, 20

\bibitem[Wang et al.(2015)]{Wang15} Wang, L., Dutton, A.~A., Stinson, G.~S., Macci\`o, A.~V., Penzo, C., Kang, X., Keller, B.~W., Wadsley, J.\ 2015, \mnras, 454, 83

 \bibitem[Wang et al.(2016)]{Wang16} Wang, L., Dutton, A.~A., Stinson, G.~S., et al.\ 2016, arXiv:1601.00967 

\bibitem[Weinberg \& Katz(2002)]{Weinberg02} Weinberg, M.~D.,
  \& Katz, N.\ 2002, \apj, 580, 627



\bibitem[Zhao et al.(2009)]{Zhao09} Zhao, D.~H., Jing, Y.~P., 
Mo, H.~J., B{\"o}rner, G.\ 2009, \apj, 707, 354 

\end{thebibliography}
\end{document}